\documentclass[twocolumn,tighten,twocolappendix]{aastex631}

\usepackage{amsmath}

\newcommand{\RE}{\mathrm{Re}}
\newcommand{\IM}{\mathrm{Im}}

\def\NP99{NP99}

\begin{document}

\title{Linear bending wave propagation in laminar and turbulent discs}

\author[0000-0002-9285-2184]{Callum W. Fairbairn}
\affiliation{Institute for Advanced Study \\
1 Einstein Drive, \\
Princeton NJ 08540, USA}




% ============================ %
\begin{abstract}
% ============================ %
Bending waves are perhaps the most fundamental and analytically tractable phenomena in warped disc dynamics. In this work we conduct 3D grid-based, numerical experiments of bending waves in laminar, viscous hydrodynamic and turbulent, weakly magnetised discs, capturing their behaviour in unprecedented detail. We clearly elucidate the theory from first principles, wherein the general Fourier-Hermite formalism can be simplified to a reduced framework which extends previous results towards locally isothermal discs. We obtain remarkable agreement with our laminar simulations wherein the tilt evolution is well described by the reduced theory, whilst higher order vertical modes should be retained for capturing the detailed disc twisting and internal velocity profiles. We then relax this laminar assumption and instead launch bending waves atop a magnetorotationally turbulent disc. Although the turbulence can be quantified with an effective $\alpha$ parameter, the bending waves behave distinctly from a classical viscous evolution and are readily disrupted when the turbulent velocity is comparable to the induced warping flows. This may have implications for the inclination damping rates induced by planet-disc interactions, the capture rate of black holes in AGN discs or the warped shapes assumed by discs in misaligned systems.
\end{abstract}



% ============================ %
\section{Introduction} 
\label{sec:intro}
% ============================ %

Distorted discs which depart from classical, circular and coplanar descriptions are now understood to be ubiquitous. In particular a diversity of observational evidence points towards the prevalence for warped discs across a range of astrophysical scales, wherein the orbital plane varies as a function of radius. For example, the periodic variability of X-ray binaries \citep{Katz_1973,KotzeCharles_2012}, the scattered light shadows \citep{MarinoEtAl_2015,DebesEtAl_2017} and non-Keplerian kinematics \citep{RosenfeldEtAl_2012,WalshEtAl_2017} detected in protoplanetary discs, and the direct observation of warped protostellar \citep{SakaiEtAl_2019} or galactic midplanes \citep{MiyoshiEtAl_1995,SkowronEtAl_2019}. Such warps are to be anticipated whenever there is a misalignment present in the system. Relevant mechanisms include the Lense-Thirring torque from a misaligned, spinning black hole \citep{BardeenPetterson_1975}, the gravitational forcing due to an inclined planet/binary companion \citep{NealonEtAl_2018, LubowOgilvie_2000} and interaction with the wider environment via stellar-flybys \citep{CuelloEtAl_2019} or infalling streamers \citep{KuffmeierEtAl_2021}. 

All of these processes crucially rely on an understanding of how warps evolve spatially and temporally in discs. This important question has motivated several studies with \cite{BardeenPetterson_1975} being the first to model the disc as a series of nested annuli with a radially varying tilt orientation. However, it was later that \cite{PapaloizouPringle_1983} realized the importance of pressure in driving oscillatory, internal flows which effectively couple together these annuli and efficiently communicate the warp. Formally their description applies to long wavelength, linear warps in viscous discs wherein the $\alpha$ parameter \citep{ShakuraSunyaev_1973} is greater than the disc aspect ratio $H/R$, leading to diffusive warp evolution. Subsequently, the (nearly) inviscid regime with $\alpha < H/R$ was considered by \cite{PapaloizouLin_1995} and \cite{LubowOgilvie_2000}, who found that long wavelength warps propagate in a wave-like fashion at half the sound speed. Theoretical efforts to extend these linear analyses towards nonlinear warp amplitudes have been significantly advanced by \cite{Ogilvie_1999} in the diffusive regime and \cite{Ogilvie_2006} in the bending wave regime. There have also been efforts to connect this dichotomy through generalised warp equations which interpolate between the diffusive and wave-like equations \citep{MartinEtAl_2019, DullemondEtAl_2022a}.

Benchmarking these theories against numerical simulation is necessary to verify the inherent analytical assumptions and test the code performance at capturing the warp evolution. Smoothed particle hydrodynamic (SPH) simulations \cite{GingoldMonaghan_1977} are often the tool of choice, owing to the lack of preferred geometry, and have provided plentiful insight into warped dynamics in the diffusive \citep[e.g.][]{LodatoPrice_2010,FacchiniEtAl_2013} and bending regimes \citep[e.g.][]{NelsonPapaloizou1999, NealonEtAl_2015}. However, the difficulty in quantifying numerical viscosity in SPH \citep[see][]{MeruBate_2012} and the fact that the resolution is tied to Lagrangian particle density could be problematic. Indeed, \cite{DrewesNixon_2021} suggest that several SPH studies, which claim to be probing wave-like behaviour, are perhaps brought closer to the diffusive regime via anomalous viscosity. Instead, they utilize a larger number of particles ($N = 10^8$) in order to achieve $\alpha \sim 0.005 < H/R$ throughout the bulk of the disc.\footnote{It should be noted that in recent years \cite{DengEtAl_2020, DengOgilvie_2022} have employed an alternative hybrid Godunov-Lagrangian \textit{Meshless Finite Mass} (MFM) scheme in \texttt{GIZMO} \citep{Hopkins_2015} to study wave-like warped discs, claiming a reduced numerical viscosity compared with SPH. Indeed with 120 million particles they attain $\alpha < 0.001$.}

This promotes grid-based codes as an alternative, to which end there have been a handful of numerical experiments making direct comparison with warped disc evolutionary theory. Notably, \cite{FragnerNelson_2010} investigated the development of warps in circumbinary discs and \cite{ArzamasskiyEtAl_2018} looked at the damping of planetary inclinations by excitation of bending waves \citep{TanakaWard_2004}. These studies begin with a brief examination of freely propagating bending waves which are primarily used as a sanity-check to benchmark their code performance. This follows in the spirit of the earlier, detailed SPH study of \cite{NelsonPapaloizou1999} (hereafter \NP99) who considered how an initially warped pulse decomposes into counter propagating bending waves. A similar SPH test was also performed more recently by \cite{NealonEtAl_2015} who used a higher number of particles compared to \NP99 and therefore achieved a lower numerical viscosity. All these experiments compared their results with a reduced, vertically integrated theory which approximates the horizontal flow fields as being low order modes, which are linear in the vertical coordinate \citep{PapaloizouLin_1995}. Whilst \cite{FragnerNelson_2010}, \cite{NealonEtAl_2015} and \cite{ArzamasskiyEtAl_2018} all claim very good agreement with this formalism, compared with the low particle number SPH study of \NP99, there is still much more that can be gleaned by revisiting such controlled experiments. Indeed, in the intervening years since \NP99, whilst there has been a plethora of SPH and grid based simulations probing a variety of phenomenological distorted disc configurations, the propagation of bending waves is usually employed as a brief test of the numerics and compressed into a cursory comparison with theory. Whilst the diffusive warp regime has received the majority of attention and been carefully compared with the appropriate theory \citep{LodatoPringle_2007, LodatoPrice_2010}, the theoretical approximations in the low viscosity regime merit closer examination in the context where the warping wavelength is not well separated from the scale height. Furthermore, the previous numerical comparisons preclude a detailed examination of the internal flow fields, which underpin laminar warp evolution theory. We will address these shortcomings fully in this work. This will nicely complement the recent study of \cite{KimmigDullemond_2024} who performed grid-based simulations of larger amplitude warps and compared with the generalised warp equations of \cite{DullemondEtAl_2022a}, applicable to longer wavelength distortions.

These previous theoretical and numerical efforts should also be caveated by the assumption of laminar flow structures. In reality discs are subject to a rich diversity of instabilities which could disrupt these smooth velocity fields and lead to turbulence --- e.g. the magnetorotational instability (MRI) \citep{BalbusHawley_1991}, the vertical shear instability (VSI) \citep{NelsonEtAl_2013} or the subcritical baroclinic instability \citep{LesurPapaloizou_2010}. Indeed, these various mechanisms can be vigorous or deactivated in different environments; leading to the relatively high turbulent states powering accreting active galactic nuclei (AGN) \citep{BeckwithEtAl_2011,FlockEtAl_2011} or the lower turbulent states inferred from observations of Class II protoplanetary discs \citep{FlahertyEtAl_2017,DullemondEtAl_2022b}. However, performing a detailed examination of warp evolution in such turbulent discs has rarely been considered. Exceptionally, \cite{SorathiaEtAl_2013a,SorathiaEtAl_2013b} performed (magneto-)hydrodynamic simulations of nonlinearly warped discs and found the presence of shocks and MRI turbulence produce significant departures from laminar theory. However, up until now, complementing these large amplitude results with the more controlled linear theory has not been examined.

To this end, we presently perform a focused study of short wavelength bending waves in both laminar and turbulent discs. This will place a renewed emphasis on a detailed comparison between theory and numerics. We exemplify that careful simulation setups are perfectly able to capture the inviscid bending waves, even for very small warp amplitudes. Furthermore, we demonstrate the clear departures from theory as turbulence disrupts the smooth evolution. Firstly, in Section \ref{sec:theory} we will expound the general linear bending wave theory and show how it can be simplified to a reduced framework for locally isothermal discs (extending the formalism of \NP99). Then in Section \ref{sec:hydro_method} we will outline the numerical method used to conduct simulations of hydrodynamic, laminar bending waves. In Section \ref{sec:laminar_results} we will examine the results from our experiments and find that with correct application of the theory we can achieve remarkably detailed agreement. Next, in Section \ref{sec:mri_method} we describe how our numerical setup is generalised towards a magnetised disc. The ensuing turbulent state is characterised in Section \ref{sec:mri}, before bending waves are input and we examine their modified propagation. We discuss our results in Section \ref{sec:discussion} before concluding remarks in Section \ref{sec:conclusion}.

% ============================ %
\section{Linear bending wave theory} 
\label{sec:theory}
% ============================ %

In this section we will elucidate the fundamental theory governing bending wave propagation in astrophysical discs. Such a detailed exposition is required for a robust comparison with the numerical studies conducted later in this paper. In Section \ref{subsec:background} we will review the steady state equilibrium for a thin disc. Then, in Section \ref{subsec:hermite} we will project the linearized fluid disturbances about this state onto a Fourier-Hermite basis which yields a set of coupled bending wave equations. These will be further simplified in Section \ref{subsec:reduced_bending} wherein we retain only the lowest order vertical modes. We will demonstrate how this connects with the previous vertically-averaged bending wave formalism and nicely extends this theory towards locally isothermal temperature structures.

% ============================ %
\subsection{Background equilibrium} 
\label{subsec:background}
% ============================ %

I will presently describe our background disc equilibrium in cylindrical coordinates $(R,\phi,z)$. Note, that these should be distinguished from the spherical coordinates denoted $(r,\theta,\phi)$ which are also used later on this paper.\footnote{We will often perform analyses as functions of the spherical radius $r$ which suits our choice of numerical grid. However, in diagnostics where the other dimensions are integrated out, $r$ can be considered to lie along the midplane with $z = 0$, where clearly $R \equiv r$ and either coordinate can be used to label the numerical value of the spherical radius.} Consider the case of a thin disc around a central star of mass $M_\star$. Then the Keplerian potential is given by
\begin{equation}
    \label{eq:phi_K}
    \Phi_\textrm{K} = -\frac{G M_\star}{\sqrt{R^2+z^2}} . 
\end{equation}
Furthermore, we define the associated Keplerian velocity in the midplane to be $\Omega_\textrm{K}(R) = \sqrt{GM_\star/R^3}$. Equivalently about some reference radius $R_0$ one can write $\Omega_\textrm{K} = \Omega_\textrm{K,0}\left(R/R_0\right)^{-3/2}$ where $\Omega_\textrm{K,0} = \sqrt{GM/R_0^3}$. The thin disc approximation allows us to expand equation \eqref{eq:phi_K} such that
\begin{equation}
    \label{eq:phi_td}
    \Phi_\textrm{K} \approx \Phi_\textrm{td} = -\Omega_\textrm{K}^2 R^2+\frac{1}{2}\Omega_\textrm{K}^2 z^2 .
\end{equation}
Throughout the remainder of this paper we will employ $\Phi_\textrm{td}$ in the equations of motion, which renders them more amenable to analytical progress compared with the point mass potential. Solving for hydrostatic equilibrium then yields the steady state density structure
\begin{equation}
    \label{eq:density}
    \rho_\textrm{td}(R,z) = \rho_0 \left(\frac{R}{R_0}\right)^{-d} \exp\left[-\frac{z^2}{2 H^2}\right]
\end{equation}
where the reference midplane density is $\rho_0$ and varies radially according to the exponent $d$. Meanwhile, $H(R) \equiv c/\Omega_\textrm{K}$ defines the disc scale height which itself depends on the sound speed profile
\begin{equation}
    \label{eq:sound_speed}
    c(R) = c_0 \left(\frac{R}{R_0}\right)^{-s/2}.
\end{equation}
This describes a power law disc temperature profile according to $T \propto c^2 \propto R^{-s}$. Therefore
\begin{equation}
    \label{eq:scale_height}
    H(R) = H_0 \left(\frac{R}{R_0}\right)^{\chi}
\end{equation}
where $H_0 = c_0/\Omega_{\textrm{K},0}$ and $\chi = (3-s)/2$. Vertically integrating the density profile yields the surface density
\begin{align}
    \sigma = \int\rho_\textrm{td} \, dz & = \sqrt{2\pi}\rho_\textrm{td}(R,0) H \nonumber\\
                            & = \sigma_0 \left(\frac{R}{R_0}\right)^{\Gamma} , 
\end{align}
where $\sigma_0 = \sqrt{2\pi}\rho_0 H_0$ and $\Gamma = \chi - d$. Now solving the radial force balance gives the equilibrium rotational profile
\begin{equation}
    \label{eq:omega}
    \Omega(R,z) = \Omega_{K}\sqrt{1-(d+s)h^2-\frac{s}{2}\left(\frac{z}{R}\right)^2} ,
\end{equation}
where $h(R) = H/R$ is the disc aspect ratio. Notice that for globally isothermal discs with $s = 0$, the disc uniformly rotates on cylinders with $\Omega_\textrm{g}(R) = \Omega_{K}\sqrt{1-d h^2}$. Otherwise, a locally isothermal temperature gradient incurs a small vertical shear flow. It is useful to expand $\Omega$ assuming $h \ll 1$ such that
\begin{equation}
    \label{eq:omega_thin_expand}
    \Omega(R,z) \approx \Omega_{\textrm{g}}(R)\left[1-\frac{s h^2}{4}\left(\frac{z}{H}\right)^2\right]
\end{equation}
where negligible errors enter at $\mathcal{O}(h^4)$. The velocity profile immediately follows from $\mathbf{u} = (0,R\Omega,0)$.

% ============================ %
\subsection{General Fourier-Hermite perturbation expansion} 
\label{subsec:hermite}
% ============================ %

The aforementioned disc equilibrium is an exact solution to the inviscid Euler equations of motion, furnished with the thin disc potential $\Phi_\textrm{td}$. Thermodynamically we assume a (locally) isothermal equation of state (EoS) linking pressure and density via $P = c^2 \rho$, thus removing the need to explicitly solve an energy equation. We now consider the net velocity and density fields to be the sum of the background plus a perturbed (primed) component --- i.e. $\mathbf{v} = \mathbf{u}+\mathbf{u}^\prime$ and $\rho = \rho_\textrm{td}+\rho^\prime$. Inserting these into the momentum and continuity equations and linearising yields
\begin{align}
    \label{eq:euler_perturbation}
    & \frac{\partial \mathbf{u^\prime}}{\partial t}+(\mathbf{u}\cdot\nabla)\mathbf{u^\prime}+(\mathbf{u^\prime\cdot\nabla})\mathbf{u} = -\nabla W - \frac{s}{R} W \hat{\mathbf{e}}_R , \\
    \label{eq:continuity_perturbation}
    & \frac{\partial \rho^\prime}{\partial t}+\nabla\cdot\left(\rho_\textrm{td}\mathbf{u}^\prime+\mathbf{u}\rho^\prime\right) = 0 , 
\end{align}
where $W = c^2 \rho^\prime/\rho_\textrm{td}$ is the pseudo-enthalpy perturbation. These equations are similar in form to those appearing in \cite{ZhangLai_2006} and \cite{LaiZhang_2008} except now we include the additional enthalpy term, proportional to $s$, in equation \eqref{eq:euler_perturbation} which accounts for the radial temperature structure. This extra contribution has recently featured in the analysis of \cite{TanakaOkada_2024} who examined the local excitation of waves launched by planets interacting with the surrounding disc. Following these studies, we make progress by Fourier decomposing equations \eqref{eq:euler_perturbation}-\eqref{eq:continuity_perturbation} azimuthally. Perturbed quantities then take the form $x^\prime(R,\phi,z) = \sum_m x_m^\prime(R,z) \exp(i m\phi)$ for mode numbers $m \geq 0$. Meanwhile, in the stratified vertical direction we can project the disturbances onto the Hermite polynomial basis \citep{OkazakiEtAl_1987, LaiZhang_2008, TanakaOkada_2024} according to 
\begin{align}
\label{eq:He_expansion}
    \begin{bmatrix}
           W_m(R,z,t) \\
           u_{R,m}^\prime(R,z,t) \\
           u_{\phi,m}^\prime(R,z,t)
    \end{bmatrix} = \sum_{n}
    \begin{bmatrix}
        W_{mn}(R,t) \\
        u_{R,mn}^\prime(R,t) \\
        u_{\phi,mn}^\prime(R,t)
    \end{bmatrix}
    \textrm{He}_n(Z)
  \end{align}
and
\begin{equation}
\label{eq:He'_expansion}
    u_{z,m}^\prime(R,z,t) = \sum_{n} u_{z,mn}^\prime(R,t)\, \frac{d \textrm{He}_n(Z)}{dZ}.
\end{equation}
Here $Z = z/H$ is the non-dimensionalised vertical coordinate and $\textrm{He}_n(Z) \equiv (-1)^n \exp(Z^2/2)d^n[\exp(-Z^2/2)]/dZ^n$ are the `probabilist's' Hermite polynomials of integer order $n$. Intuitively, the order $n$ denotes the number of vertical nodes contained in that mode. These polynomials satisfy a variety of useful properties \citep[cf.][]{Ogilvie_2008}, namely the recurrence relations
\begin{align}
    \label{eq:orthogonality_condition}
    & \frac{d \textrm{He}_{n}(Z)}{dZ} = n \textrm{He}_{n-1}(Z) , \\
    & Z \textrm{He}_{n}(Z) = \textrm{He}_{n+1}(Z)+ n \textrm{He}_{n-1}(Z) ,
\end{align}
and the orthogonality condition 
\begin{equation}
    \label{eq:orthogonal}
    \int_{-\infty}^{\infty}\textrm{He}_j(Z)\textrm{He}_k(Z) \exp(-Z^2/2) dZ = \sqrt{2\pi}j!\delta_{jk}.
\end{equation}
This Fourier-Hermite expansion is inserted into equations \eqref{eq:euler_perturbation}-\eqref{eq:continuity_perturbation} and the orthogonal inner product given by \eqref{eq:orthogonal} projects onto the coefficients. In this process we use the expanded background rotational profile given by equation \eqref{eq:omega_thin_expand}, wherein the locally isothermal vertical shear dependence is easily handled since $Z^2 = \textrm{He}_2+\textrm{He}_0$. This yields a hierarchy of equations
\begin{align}
    %-----------------------------
    \label{eq:fourier_hermite_uR} 
    %-----------------------------
    &\frac{\partial u_{R,n}^\prime}{\partial t} + i m \Omega_\textrm{g} u_{R,n}^\prime - 2 \Omega_\textrm{g} u_{\phi,n}^\prime \nonumber\\
    &- i m \Omega_\textrm{g} h^2 \frac{s}{4} \left[u_{R,n-2}^\prime+(2n+1)u_{R,n}^\prime+(n+1)(n+2)u_{R,n+2}^\prime\right] \nonumber\\
    & +2 \Omega_\textrm{g} h^2 \frac{s}{4}\left[u_{\phi,n-2}^\prime+(2n+1)u_{\phi,n}^\prime+(n+1)(n+2)u_{\phi,n+2}^\prime\right] \nonumber\\
    & = -\frac{\partial W_{n}}{\partial R}+\frac{n\chi}{R}W_{n}+\frac{(n+1)(n+2)\chi}{R}W_{n+2}-\frac{s}{R}W_{n} , \\
    %-----------------------------
    \label{eq:fourier_hermite_uphi} 
    %-----------------------------
    &\frac{\partial u_{\phi,n}^\prime}{\partial t} + im\Omega_\textrm{g} u_{\phi,n}^\prime+\frac{1}{R}\frac{d(R^2\Omega_\textrm{g})}{dR}u_{R,n}^\prime \nonumber\\
    &-im\Omega_\textrm{g}h^2\frac{s}{4}\left[u_{\phi,n-2}^\prime+(2n+1)u_{\phi,n}^\prime+(n+1)(n+2)u_{\phi,n+2}^\prime\right] \nonumber\\
    &-\frac{s}{4}h^2 R \frac{\partial\Omega_\textrm{g}}{\partial R}\left[u_{R,n-2}^\prime+(2n+1)u_{R,n}^\prime+(n+1)(n+2)u_{R,n+2}^\prime\right] \nonumber\\
    &-2 \Omega_\textrm{g} h \frac{s}{4}\left[n u_{z,n}^\prime + (n+1)(n+2)u_{z,n+2}^\prime\right] = -\frac{i m}{R} W_{n} , \\
    %-----------------------------
    \label{eq:fourier_hermite_uz} 
    %-----------------------------
    & \frac{\partial u_{z,n}^\prime}{\partial t} + im\Omega_\textrm{g} u_{z,n}^\prime -i m \Omega_\textrm{g} h^2 \frac{s}{4} \biggl[\frac{(n-2)}{n}u_{z,n-2}^\prime \nonumber\\
    & +(2n-1)u_{z,n}^\prime+(n+1)(n+2)u_{z,n+2}^\prime \biggr] = -\frac{W_{n}}{H} , \\
    %-----------------------------
    \label{eq:fourier_hermite_W} 
    %-----------------------------
    &\frac{\partial W_{n}}{\partial t} + im\Omega_\textrm{g} W_{n} \nonumber\\
    &-i m \Omega_\textrm{g}h^2\frac{s}{4}\left[W_{n-2}+(2n+1)W_{n}+(n+1)(n+2)W_{n+2}\right] \nonumber\\
    &=-c^2\biggl\{ \frac{d}{dR}\left[\ln(\sigma R)+\frac{n\chi}{R}\right]u_{R,n}^\prime+\frac{d u_{R,n}^\prime}{dR}+\frac{\chi}{R}u_{R,n-2}^\prime \nonumber\\
    &\frac{im}{R}u_{\phi,n}^\prime-\frac{n u_{z,n}^\prime}{H}\biggr\} .
\end{align}
Here we have dropped the subscript $m$ for notational convenience. This causes no ambiguity since each set of equations, for a given azimuthal mode, are decoupled in this linear analysis. Meanwhile, we see that the vertical modes are coupled owing to background gradients in the vertical and radial disc structure. This coupling allows for communication between modes of order $n$ and $n\pm2$, such that only modes of the same parity talk to each other. When the disc is globally isothermal ($s=0$) we see that these equations reduce substantially, collapsing to the previous description contained within equations (20)-(23) of \cite{ZhangLai_2006}, for example. It should be noted that so far we have neglected viscosity in deriving these equations. However, in appendix \ref{app:viscous} we will discuss how these might be modified in the presence of a non-zero $\alpha$ prescription.

Throughout this paper, we will solve these partial differential equations using the \textit{Method of Lines} technique. This approach discretizes over the remaining radial dimension and instead solves an ordinary differential equation at each grid point using an `RK45' solver. We will adopt 100 radial grid points and equip these with fixed boundary conditions at the inner and outer disc edges. The initial conditions are obtained by projecting the full perturbation onto each Fourier-Hermite basis combination. Formally, the equation set couples vertical modes in an infinite ladder, but in practice only a finite combination of low modal number contribute significantly. We will specify the included modes and discuss their importance in the upcoming results sections.

% ============================ %
\subsection{Reduced bending wave description} 
\label{subsec:reduced_bending}
% ============================ %

These coupled Fourier-Hermite equations give a comprehensive description of linear disc perturbations. However, we might also seek a reduced formalism which preserves much of the key qualitative behaviour, similar to that described in \NP99. In particular we are interested in bending waves which are dominated by $m=1$ and $n=1$ perturbations \citep[c.f.][]{PapaloizouLin_1995}. Indeed, infinitesimal rigid tilts to flat discs induce Eulerian perturbations for the horizontal velocity components which are linear in the coordinate $z = \textrm{He}_{1}$. Meanwhile, the vertical velocity component is constant ($\propto \textrm{He}_{0}$) in $z$ at leading order in the tilt amplitude. Presently we restrict our attention to this single modal combination and therefore all coupling terms in equations \eqref{eq:fourier_hermite_uR}-\eqref{eq:fourier_hermite_W} disappear. Furthermore, we will drop the locally isothermal terms which are multiplied by $h^2$. However, it is important to retain the $s h$ term in equation \eqref{eq:fourier_hermite_uphi} which actually counteracts the effects of the $s$ contribution in equation \eqref{eq:fourier_hermite_uR}. Indeed, for a slowly evolving perturbation (with characteristic timescale $\gg$ dynamical timescale), the dominant balance in equation \eqref{eq:fourier_hermite_uz} implies $\mathcal{O}(W)\sim \mathcal{O}(h u_{z}^\prime)$. If we now make the identifications
\begin{align}
    \label{eq:reduced_hermite_relations}
    & q_R(R,t) = u_{R,1}^\prime/H ,  \\
    & q_\phi(R,t)  = u_{\phi,1}^\prime/H , \\
    & v_z(R,t)  = u_{z,1}^\prime , \\
    & g(R,t)  = i \frac{W_1}{R\Omega_\textrm{g}^2 H},
\end{align}
and insert these into the simplified Fourier-Hermite equations, after some manipulation one arrives at
\begin{align}
    \label{eqn:NP_qr}
    & \frac{\partial q_R}{\partial t} +i\Omega_\textrm{g} q_R-2\Omega_\textrm{g} q_\phi = i\frac{\partial (R\Omega_\textrm{g}^2 g)}{\partial R} +i s \Omega_\textrm{g}^2g , \\
    \label{eqn:NP_qphi}
    & \frac{\partial q_\phi}{\partial t} +i\Omega_\textrm{g} q_\phi+\frac{q_R}{R}\frac{\partial(R^2\Omega_\textrm{g})}{\partial R}-\frac{s}{2 R}\Omega_\textrm{g}v_z = -\Omega_\textrm{g}^2 g , \\
    \label{eqn:NP_vz}
    & \frac{\partial v_z}{\partial t}+i\Omega_\textrm{g} v_z = i R\Omega_\textrm{g}^2 g , \\
    \label{eqn:NP_delta}
    & \mathcal{J} R\Omega_\textrm{g}^2\left(\frac{\partial g}{\partial t}+i\Omega_\textrm{g} g\right) = -\frac{i}{R}\frac{\partial(R\mu q_R)}{\partial R}+\frac{\mu q_\phi}{R}+i\sigma v_z ,
\end{align}
where the vertical moments are defined as $\mathcal{J} = \mu/c^2$ and $\mu = \sigma H^2$. In the globally isothermal case these equations are identical to the vertically averaged linear bending wave theory encapsulated by equations (8)-(9) of \NP99. Here we have supplemented these with two additional terms proportional to $s$ which generalise their applicability towards locally isothermal discs. As was discussed by \NP99, these equations admit the trivial solution corresponding to a uniform, rigid disc inclination of tilt amplitude $g$ for which $q_R = -\Omega g$, $q_\phi=-i g [d(R\Omega)/dR]$ and $v_z = R \Omega g$. Of course, for a point mass spherical potential, such rotations are obvious symmetries of the system. Although our derivation has assumed an expanded thin disc potential which introduces a `preferred' vertical direction, for infinitesimal tilts of thin discs they look the same. We should also recognise that in this reduced approach, by restricting our attention only to the odd $n=1$ mode, we remove the even azimuthal vertical shearing term which is intrinsic to locally isothermal disc equilibria, as can be seen in equations \eqref{eq:omega}-\eqref{eq:omega_thin_expand}. If one cares about the fine details of the internal flows then a better approach is to use the full Fourier-Hermite method developed in the previous section. Finally, since these reduced equations are simply a restriction of the more general Fourier-Hermite modal analysis, we will numerically solve these using the same Method of Lines technique discussed at the end of Section \ref{subsec:hermite}, where now we only consider the single modal combination $m=1$ and $n=1$. In appendix \ref{app:long_wavelengths} we further simplify this reduced framework, in the asymptotically long bending wavelength limit, to connect with the classic, large scale warping theory developed by \cite{LubowOgilvie_2000}.
% ============================ %
\section{Hydrodynamic numerical method} 
\label{sec:hydro_method}
% ============================ %

In order to validate this linear theory we will now appeal to numerical simulation. In Section \ref{subsec:hydro_eqns} we outline the numerical equations being solved, before describing how to set up our initial conditions in Section \ref{subsec:hydro_ics}. Next we outline the numerical parameters adopted in Section \ref{subsec:hydro_params} before finally discussing the resolution of our simulations in Section \ref{subsec:resolution}.

% ---------------------------- %
\subsection{Equations being solved }
\label{subsec:hydro_eqns}
% ---------------------------- %

We will first conduct 3D, hydrodynamic simulations using the finite-volume, grid-based Godunov astrophysical fluid code \texttt{Athena++} \citep{StoneEtAl_2019}. The general conservative equations being solved are
\begin{align}
    \label{eqn:athena_continuity}
    & \frac{\partial \rho}{\partial t}+\nabla\cdot\left[ \rho\mathbf{v} \right] = 0 , \\
    \label{eqn:athena_momentum}
    & \frac{\partial (\rho \mathbf{v})}{\partial t} + \nabla\cdot\left[ \rho \mathbf{v}\mathbf{v}+P\mathbf{I}-\mathbf{\Pi}\right]+\rho\nabla\Phi_\textrm{td} = 0, \\
    \label{eqn:athena_eos}
    & P = c(R)^2 \, \rho,
\end{align}
where $\rho$ is the density, $\mathbf{v}$ is the velocity vector, $c$ is the locally isothermal sound speed and $P$ is the pressure combined with the isotropic identity tensor $\mathbf{I}$. Meanwhile, $\mathbf{\Pi}$ is the full, shear viscous stress tensor given by
\begin{equation}
    \label{eqn:athena_viscous}
    \mathbf{\Pi} = \rho\nu\left(\nabla\mathbf{v}+\nabla\mathbf{v}^{T}-\frac{2}{3}\nabla\cdot\mathbf{v}\mathbf{I}\right).
\end{equation}
We set the kinematic shear viscosity in accordance with the standard $\alpha$ prescription \citep{ShakuraSunyaev_1973} so that $\nu = \alpha c H$ for an assumed turbulent viscosity acting on velocity scales a fraction of the sound speed and eddy length scales a fraction of the scale height $H$. Whilst we focus our attention on `inviscid' discs in Section \ref{sec:laminar_results}, we do examine viscous runs in Section \ref{subsec:turbulent_bending} and in appendix \ref{app:viscous}. Note that although these `inviscid' runs do not include an explicit physical viscosity, there will of course be some inherent numerical viscosity. However, as we will see in Section \ref{subsec:sloshing_flows}, the remarkable agreement of the inviscid theory with our $\alpha = 0.0$ numerical experiments constrains the numerical viscosity to be very small. In fact, numerical tests with a physical $\alpha = 0.005$ (see appendix \ref{app:viscous}) or even $\alpha = 0.001$ (not shown) produce notable changes in the flow structure compared with the inviscid theory. Thus, our underlying numerical $\alpha$ is certainly thought to be less than this for the resolutions employed in this paper.

Also notice, that unless stated otherwise, we are implementing the thin disc potential source term, rather than the inbuilt point mass potential. This enables us to benchmark the theoretical predictions to a high level of precision without conflating differences induced by numerical effects and analytical approximations. We will revisit the effect of using the point mass potential instead in Section \ref{subsec:sloshing_flows}.

For globally isothermal simulations the code has native support for a reduced thermodynamic space where no energy equation is solved. Meanwhile, for locally isothermal setups a redundant energy equation is still solved and we manually enforce the spatially dependent pressure-density relationship at each timestep. This is equivalent to the Newtonian cooling method used in other studies \citep[e.g.][]{ZhuEtAl_2015,ArzamasskiyEtAl_2018,SvanbergEtAl_2022} in the limit of instantaneous temperature relaxation. 

% ---------------------------- %
\subsection{Initial conditions}
\label{subsec:hydro_ics}
% ---------------------------- %

Since warped discs are characterised by a series of nested, tilted orbits which exhibit a radial variation in the orientation of fluid annuli (cf. \cite{OgilvieLatter2013a} Fig.~1), the initial conditions and subsequent analysis are most naturally formulated in terms of spherical coordinates\footnote{Note however, that the flows are still not aligned with the grid and numerical precession of tilted orbits can affect results for low resolution runs \citep[see][]{KimmigDullemond_2024}.} $(r,\theta,\phi)$. We first consider the unperturbed, flat disc equilibrium described in Section \ref{subsec:background} where the density and azimuthal velocity structures are given by equations \eqref{eq:density} and \eqref{eq:omega} respectively. Now, in order to initialise the warped profile we will rotate this flat equilibrium towards the target distorted state which is described by the radial warping profile
\begin{equation}
    \label{eq:target_profile}
    \beta_\textrm{init}(r) = \beta_\textrm{amp}\exp[-(r-r_\textrm{mid})^2].
\end{equation}
Here, $\beta_\textrm{amp}$ defines the amplitude of the tilt and $r_\textrm{mid}$ governs the radial location of maximal tilting. Then each initialisation cell $\mathbf{x}$, is mapped to the unwarped state $\mathbf{x}_\textrm{f}$ via the linear transformation
\begin{equation}
    \mathbf{x}_\textrm{f} = \mathbf{R}\cdot\mathbf{x} \quad \textrm{where} \quad 
    \mathbf{R} = \begin{pmatrix}
    \cos\beta & 0 &-\sin\beta \\
    0 & 1 & 0 \\
    \sin\beta & 0 & \cos\beta
    \end{pmatrix}.
\end{equation}
This allows us to assign the warped primitive variables according to their values in the flat disc state via
\begin{align}
    \label{eq:ic_rotate_rho}
    & \rho(\mathbf{x}) = \rho_\textrm{f}(\mathbf{x}_\textrm{f}) , \\
    \label{eq:ic_rotate_u}
    & \mathbf{u}(\mathbf{x}) = \mathbf{R}^\textrm{T}\cdot\mathbf{u}_\textrm{f}(\mathbf{x}_\textrm{f}) ,
\end{align}
where the transpose operator $\mathbf{R}^\textrm{T}$ ensures the velocity vector components are also rotated accordingly.

% ---------------------------- %
\subsection{Numerical parameters}
\label{subsec:hydro_params}
% ---------------------------- %

To enact these initial conditions we must choose some standard parameter choices for our simulations. We are free to set the code reference length, mass and time-scales such that $R_0 = \Omega_{\textrm{K},0} = \rho_0 = 1.0$. In all runs we will adopt the power law density profile with $d = 2$. We will then focus our study on two illustrative temperature structures --- globally isothermal runs denoted \texttt{giso} with $s=0$ and locally isothermal runs denoted \texttt{liso} with $s = 1$. For \texttt{giso} we adopt the constant sound speed $c_0 = 0.035$, whilst for \texttt{liso} we take $c_0 = 0.08$. 

In both simulations, the domain covers $r = [1,10]$ and the full azimuthal range $\phi = [0,2\pi]$. For \texttt{giso} the aspect ratio is flared with $h \propto R^{1/2}$, whilst \texttt{liso} maintains a constant $h$. During testing we find that this different flaring behaviour requires the adoption of a wider polar domain for \texttt{liso} to prevent the thicker inner disc from interacting with the vertical boundaries. Therefore, we set $\theta = [\pi/2 \pm 0.525]$ and $\theta = [\pi/2 \pm 0.35]$ for \texttt{liso} and \texttt{giso} runs respectively. For reference, the flared structure of \texttt{giso} means that the polar domain spans $\pm 10H$ at $r = 1$, $\pm 4.5 H$ at $r = 5$ and $\pm 3.1 H$ at the outer boundary $r=10$. Meanwhile, for \texttt{liso} the polar domain uniformly spans $\pm 6.6 H$ across all radii. This domain is also equipped with appropriate boundary conditions. The azimuthal direction obviously admits periodic boundaries in all variables. Meanwhile the primitive variables are all simply set to the fixed, initial conditions in the radial and polar directions. 

Finally, the warping midpoint is always chosen to lie at $r_\textrm{mid} = 5$. Note that at this location the scale height $H_\textrm{mid} = 0.4$ for both the \texttt{giso} and \texttt{liso} runs. We will then warp the disc with different amplitudes, $\beta_\textrm{amp} = [0.00175,0.0175,0.035,0.07]$ which progress from the firmly linear regime towards weakly nonlinear tilts. In more familiar units this corresponds to tilts of $[0.1,1.0,2.0,4.0]$ degrees. 

% ---------------------------- %
\subsection{Resolution}
\label{subsec:resolution}
% ---------------------------- %

Capturing the subtle internal fluid flows induced by such small amplitude warps is a considerable numerical challenge which requires sufficient resolution. Indeed, the previous detailed SPH study of \NP99 was computationally limited by a relatively low particle number and therefore considered larger tilts of $> 5$ degrees, which are in fact already engaging nonlinear effects. In order to resolve the linear sloshing flows we therefore desire the cell size to be some small fraction of the scale height. For the results in Section \ref{sec:laminar_results} we take $N_r = 512$ and $N_\phi = 1024$ for the number of radial and azimuthal cells. Meanwhile, in the polar direction we choose $N_\theta = 288$ for \texttt{giso} and $N_\theta = 432$ for \texttt{liso}, which scales with the vertical domain size so the angular resolution remains the same. At the midpoint location $r_\textrm{mid} = 5$ and $\theta = \pi/2$ this corresponds to $(\Delta r, r_\textrm{mid}\Delta\theta, r_\textrm{mid}\Delta\phi) \approx (0.08,0.04,0.03)H_\textrm{mid}$ which well captures the characteristic scale height and radial warping lengths intrinsic to the problem. We will show in Section \ref{sec:laminar_results} that this resolution is capable of capturing the bending wave at an astonishing level of detail. Whilst we do not present a comprehensive resolution study in this paper, we have performed a range of tests with different $(N_r,N_\theta,N_\phi)$ and find that even for half resolution runs the bending wave propagation is largely the same with only subtle deviations in the detailed flow fields (see appendix \ref{app:lower_res} for more details).

% ============================ %
\section{Laminar bending wave results} 
\label{sec:laminar_results}
% ============================ %

In this section we perform a detailed comparison of our hydro simulations with the theoretical expectations outlined in Section \ref{sec:theory}. Firstly, in Section \ref{subsec:comparison_diagnostics} we will describe how key warping diagnostics should be extracted and compared between simulation and theory. In Section \ref{subsec:warp_propagation} we will examine the decomposition of the warped initial condition into a pair of inwards and outwards propagating pulses. Then in Section \ref{subsec:sloshing_flows} we will look more closely at the internal flow fields generated by the bending wave. Finally we will consider the effects of weak nonlinearities in Section \ref{subsec:weak_nonlinearities}. For the effects of viscosity, one is referred to appendix \ref{app:viscous}.

% ........................................ %
\subsection{Comparison diagnostics}
\label{subsec:comparison_diagnostics}
% ........................................ %

In order to characterise the bending wave propagation we must first formalise the relevant diagnostics to be compared between our simulations and theory. In our numerical runs the most natural way to describe the warped geometry of the disc is through the integrated angular momentum in each radial shell \citep[e.g.][]{FragnerNelson_2010,ArzamasskiyEtAl_2018,KimmigDullemond_2024},
\begin{equation}
    \label{eq:int_ang_mom}
    \mathbf{L}(r) = \iint (\rho\mathbf{r}\times\mathbf{v}) r^2 \sin{\theta} \,d\theta\,d\phi.
\end{equation}
This is projected onto fixed Cartesian coordinates such that $\mathbf{L} = (L_x,L_y,L_z)$. The directional information of this vector encodes the disc tilt,
\begin{equation}
    \label{eq:tilt}
    \beta(r) = \arccos\left(\frac{L_z}{\left|\mathbf{L}\right|}\right) ,
\end{equation}
and the twist or precession angle,
\begin{equation}
    \label{eq:twist}
    \gamma(r) = \arctan\left(\frac{L_y}{L_x}\right).
\end{equation}
Meanwhile, using the cylindrical Fourier-Hermite solution we can superimpose all the perturbed modes and the background equilibrium, thereby reconstructing the full disc density and velocity structures. This can be interpolated onto a spherical grid and similarly integrated to yield $\mathbf{L}(r)$. A simplified analytical implementation of this method can also be applied to the reduced bending wave formalism. Indeed, whilst previous studies often erroneously identify $\beta = |g|$ \cite[e.g.][]{FragnerNelson_2010,ArzamasskiyEtAl_2018} this is not strictly correct. Although this is true for a rigidly tilted disc with $g=\textrm{constant}$, the bending wave induces sloshing flows which complicate the picture for time-dependent warps. For this  $n=1$, $m=1$ perturbation the net velocity is reconstructed as $\mathbf{v} = \mathbf{u} + \RE[(q_R z,q_\phi z, v_z)\exp(i\phi)]$ whilst the density is $\rho = \rho_\textrm{td}-\RE [i\rho_\textrm{td} R \Omega_\textrm{g}^2 g \exp(i\phi) z/c^2]$. Then for a thin disc we can approximate the integral over spherical shells as a vertical integral over cylindrical annuli such that
\begin{equation}
    \mathbf{L}(R) \approx \iint (\rho\mathbf{r}\times\mathbf{v}) R d\phi dz .
\end{equation}
The integrand is projected onto Cartesian components and we retain terms which are even about the midplane. Recalling the definition for the vertical moments $\sigma = \int \rho_\textrm{td}\, dz$ and $\mu = \int \rho_\textrm{td} z^2\, dz = \sigma H^2$ we eventually arrive at
\begin{align}
\label{eq:Lx_reduced}
    & L_x = \pi R \sigma R^2 \left[-h^2 \RE(q_\phi)- \frac{\IM(v_z)}{R}+h^2\IM(q_R)-\Omega_\textrm{g}\IM(g) \right], \\
\label{eq:Ly_reduced}
    & L_y = \pi R \sigma R^2 \left[ h^2 \IM(q_\phi)-\frac{\RE(v_z)}{R}+h^2 \RE(q_R)-\Omega_\textrm{g}\RE(g) \right] , \\
\label{eq:Lz_reduced}
    & L_z = 2 \pi R \sigma R^2 \Omega_\textrm{g} \left[1 -\frac{1}{2\Omega_\textrm{g}} \left(\RE(g)\IM(q_\phi)-\IM(g)\RE(q_\phi)\right)\right].
\end{align}
Dropping the $h^2$ contributions, one can show that to leading order,
\begin{equation}
    \label{eq:reduced_tilt_twist}
    \beta = \frac{1}{2}\left|g+\frac{v_z}{R \Omega_\textrm{g}}\right| \quad \textrm{and} \quad \gamma =  \textrm{arg}\left[\,\overline{{i\left(g+\frac{v_z}{R \Omega_\textrm{g}}\right)}}\,\right].
\end{equation}
Necessarily, this correctly reduces to $\beta = |g|$ for the rigid tilt solution with $v_z = R \Omega_\textrm{g}g$ . Otherwise, one must consider the average of these two perturbative effects. 

% ........................................ %
\subsection{Warp propagation}
\label{subsec:warp_propagation}
% ........................................ %

% -.-.-.-.-.-.-.-.-.-.-.-.-.-.-.-.- %
\begin{figure*}
    \centering
    \includegraphics[width=\textwidth]{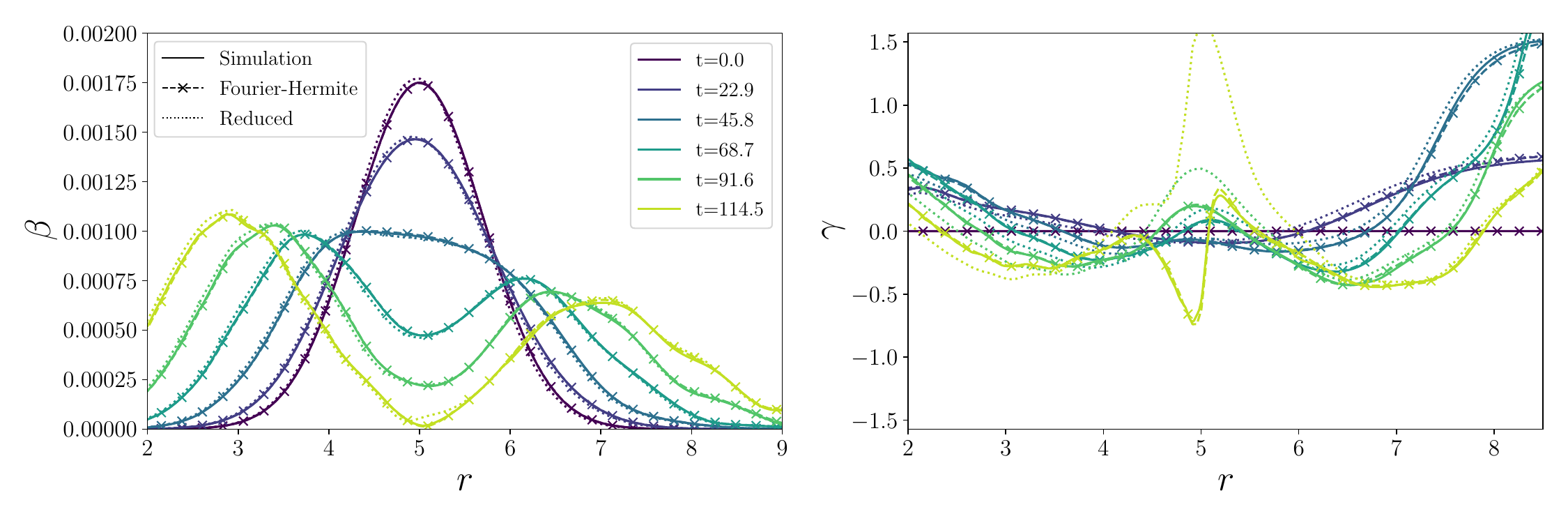}
    \caption{Comparing the warp propagation between simulation and theory for the globally isothermal case. \textit{Left panel}: the tilt evolution $\beta$. \textrm{Right panel}: the twist evolution $\gamma$. The solid lines denote the simulation results, the dashed-crossed lines show the Fourier-Hermite predictions when including $m=1$ and $n=[1,3,5]$ modes and the dotted lines show the results from the reduced bending wave theory. The different colours denote the progression of time in units of $\Omega_0^{-1}$.}
    \label{fig:hydro_linear_comparison}
\end{figure*}
% -.-.-.-.-.-.-.-.-.-.-.-.-.-.-.-.- %

The most obvious benchmark of the code-theory performance is their efficacy in capturing the bending wave propagation over time. Whilst this standard test has been qualitatively discussed by \NP99, \cite{FragnerNelson_2010} and \cite{ArzamasskiyEtAl_2018}, we will examine this in renewed detail and thereby shed light on previous discrepancies.

In Fig.~\ref{fig:hydro_linear_comparison}, we first focus on the warp propagation in our globally isothermal setup. In the left and right panels we plot the radial tilt and twist profiles respectively. To ensure we are operating firmly within the linear perturbative regime, we presently consider the lowest tilt amplitude, $\beta_\textrm{amp} = 0.00175$, such that $\beta \ll H/R$. This ensures that all Eulerian perturbations are small relative to the background equilibrium state and validates the comparison with the theory expounded in Section \ref{sec:theory}.\footnote{We will see later in Section \ref{subsec:weak_nonlinearities} that for the larger tilt amplitude runs, where $\beta \sim H/R$, the linear theory still predicts the overall tilt evolution surprisingly well, although the detailed internal flow structures begins to deviate.} The solid lines then plot the evolution for the \texttt{giso\_0.00175} run where the initial Gaussian bump in $\beta$ clearly decomposes into two inwards and outwards propagating bending waves. During this time we also see a strong development of twist which is often overlooked. We run the simulation for a bending wave crossing time, at which point interaction with the radial boundaries contaminate the subsequent results. These results are over-plotted with dashed-crossed lines which show the result from the Fourier-Hermite solution. Here we have incorporated the dominant $m=1$ azimuthal mode but considered a number of odd parity vertical modes $n = [1,3,5]$. We find that this gives remarkable agreement across all times considered, practically overlapping the simulation results in both the tilt and twist quantities. Finally, the dotted lines plot the $m=1$, $n=1$ reduced bending wave formalism, with the tilt and twist computed according to equations \eqref{eq:reduced_tilt_twist}. These also capture the tilt evolution very well with only small discrepancies. Although, it should be noted that if $|g|$ is used as the tilt diagnostic instead, there are more prominent errors as discussed in Section \ref{subsec:comparison_diagnostics}. This presumably leads to some of the small discrepancies found in Fig.~2 of \cite{ArzamasskiyEtAl_2018}. Meanwhile, the twist is far less successfully modelled by this reduced theory. Although it broadly follows the numerical and Fourier-Hermite profile at early times, it soon exhibits notable differences. In particular, these departures become most exaggerated near $r_\textrm{mid}$ at later times when the tilt here becomes small. This underlines the need to include higher order vertical modes in order to capture the fine structure of the twist evolution. The development of this $\gamma$ has also been observed in the recent numerical simulations of wave-like warps by \cite{KimmigDullemond_2024} who remark that their 1D model does not predict such twisting. They speculate that this is a physical manifestation of nonlinearities associated with their larger warp amplitudes, and filtered out by the linearization process. However, here we see that twist evolution is a fundamental prediction of the linear theory also. Indeed, the simplified form of the 1D warp equations used by \cite{KimmigDullemond_2024} were originally derived in the bending-wave regime by \cite{LubowOgilvie_2000}. This derivation notes a key requirement that the warping radial length scale should be much longer that the characteristic scale height. For completeness, in appendix \ref{app:long_wavelengths} we will formally demonstrate the connection of our general linearization with this long bending wavelength limit. However, this asymptotic scale separation is not cleanly satisfied in our simulations or those of \cite{KimmigDullemond_2024} and hence some growth of twist should be expected.

% -.-.-.-.-.-.-.-.-.-.-.-.-.-.-.-.- %
\begin{figure*}
    \centering
    \includegraphics[width=\textwidth]{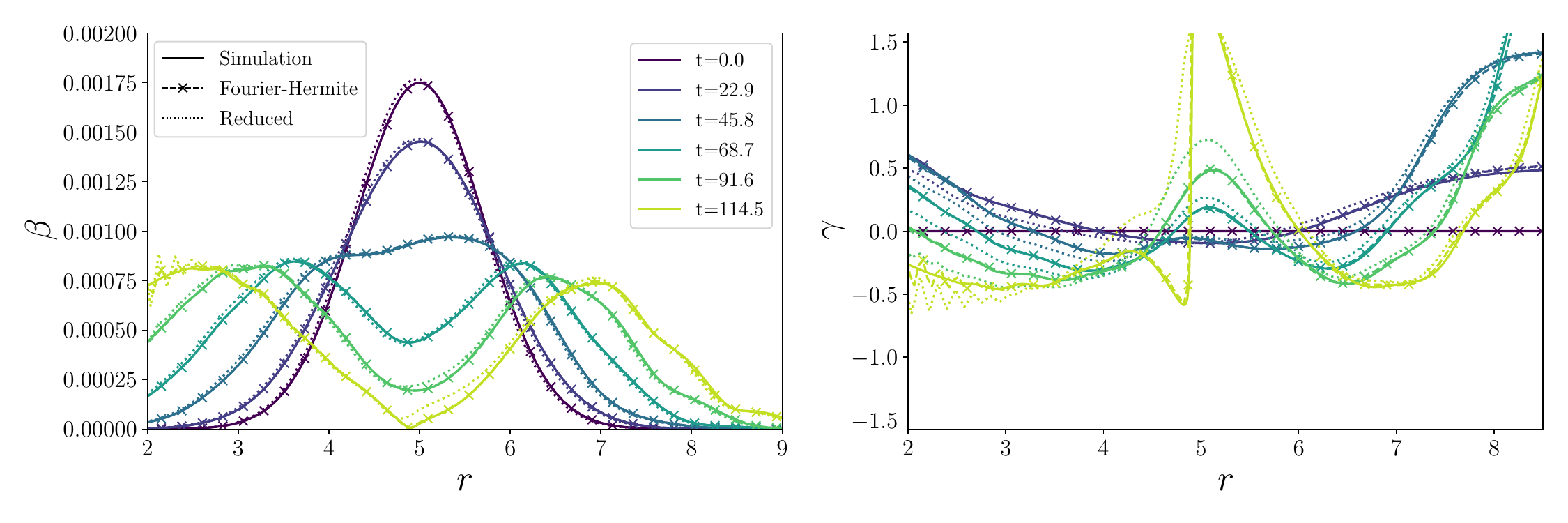}
    \caption{Same as for Fig.~\ref{fig:hydro_linear_comparison} but for the locally isothermal case.}
    \label{fig:iso_hydro_linear_comparison}
\end{figure*}
% -.-.-.-.-.-.-.-.-.-.-.-.-.-.-.-.- %

We also perform the same comparison for the locally isothermal run \texttt{liso\_0.00175} as shown in Fig.~\ref{fig:iso_hydro_linear_comparison}. In this case the pulse exhibits a modified propagation behaviour due to the radially decreasing temperature profile. Since bending waves roughly propagate at half the sound speed \citep{PapaloizouLin_1995}, the inner pulse travels faster than the outer pulse. Therefore the inner pulse reaches the boundary more quickly compared to the globally isothermal run, whilst the outer bump steepens as it slows down and `piles up' misaligned angular momentum. Furthermore, the comparison with theory tells a similar story to the globally isothermal study, with the simulation tilt excellently traced by the Fourier-Hermite and reduced bending solutions. Once again the twist behaviour is in fantastic concordance with the Fourier-Hermite method, which outperforms the reduced equations that show larger deviations. We only begin to see some spurious oscillations entering from the edges as the pulse is reflected off the boundary.

% ........................................ %
\subsection{Sloshing flows}
\label{subsec:sloshing_flows}
% ........................................ %

The previous section characterises the warp propagation via the $\beta$ and $\gamma$ diagnostics which integrate perturbations over spherical shells. However, this is a somewhat coarse-grained perspective and we can obtain a lot more information by looking at the detailed vertical structure of the perturbed flow. Despite the longstanding appreciation that internal flow dynamics play a crucial role in communicating warps \citep[e.g.][]{PapaloizouPringle_1983,Ogilvie_1999, LubowOgilvie_2000}, the extraction of these velocity perturbations from global simulations and comparison with theoretical expectations have not been sufficiently addressed. 

% -.-.-.-.-.-.-.-.-.-.-.-.-.-.-.-.- %
\begin{figure*}
    \centering
    \includegraphics[width=\textwidth]{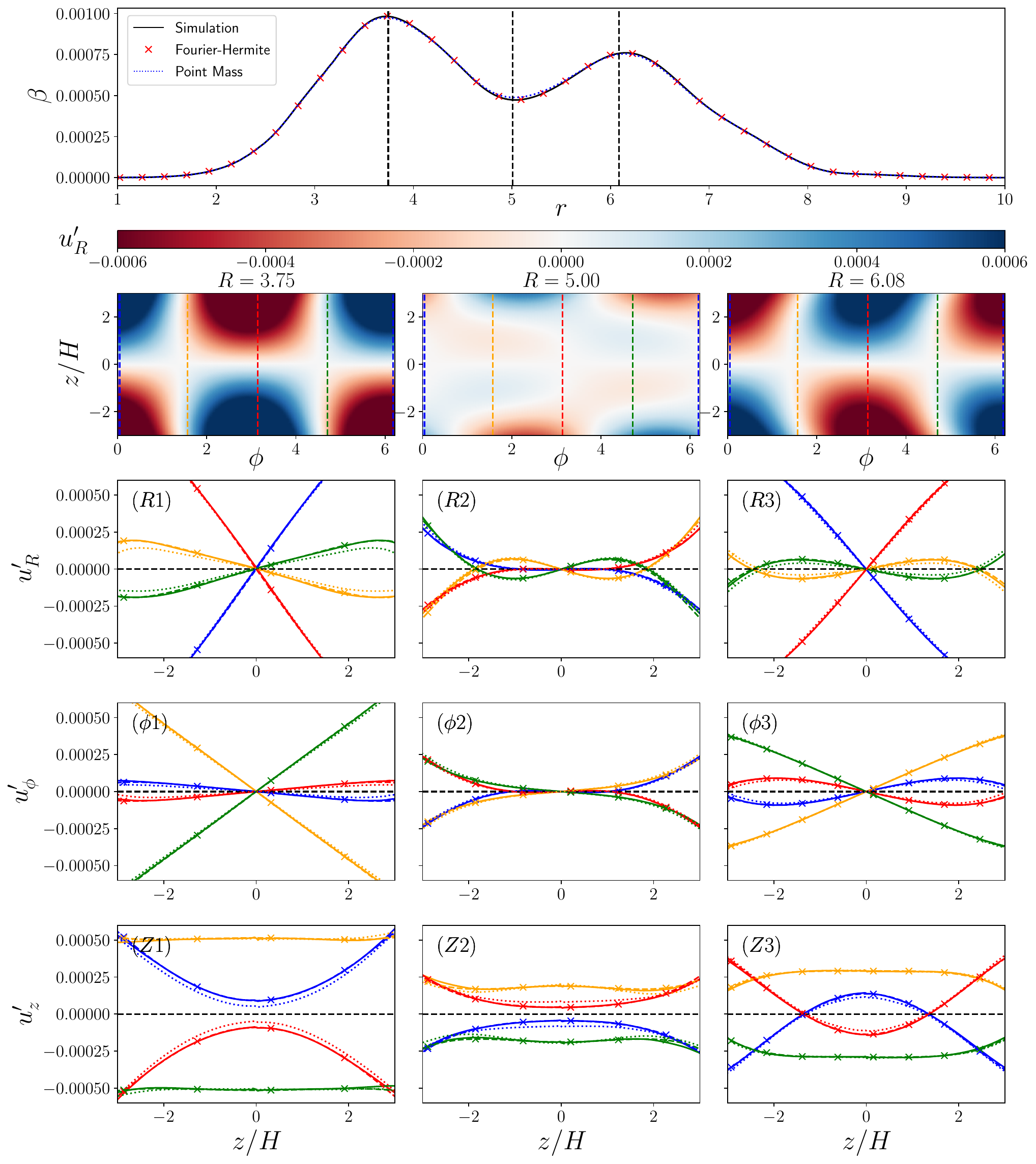}
    \caption{Globally isothermal perturbed flow fields. \textrm{Upper panel}: the tilt profile is shown for the simulation (black solid), Fourier-Hermite theory (red cross) and the point mass potential simulation (blue dots) at $t = 68.7\Omega_0^{-1}$. Dashed vertical lines denote three cylindrical radial slices at $R = [3.75,5.0,6.08]$. Azimuthal-vertical slices through these locations are shown below in the $u_R^\prime$ simulation heat maps. Four azimuthal slices are overplotted as dashed blue, yellow, red and green lines at $\phi = [0,\pi/2,\pi,3\pi/2]$. Correspondingly coloured vertical slices at each azimuthal location are shown in the panels below for $u_R^\prime$, $u_\phi^\prime$ and $u_z^\prime$. Here, $(R,\phi,Z)$ label the velocity components whilst $(1,2,3)$ denote the three radial positions. Again solid, dash-crossed and dotted lines correspond to the simulation, theory and point mass run results respectively.}
    \label{fig:global_iso_hydro_hermite_panels_0.00175}
\end{figure*}
% -.-.-.-.-.-.-.-.-.-.-.-.-.-.-.-.- %

% -.-.-.-.-.-.-.-.-.-.-.-.-.-.-.-.- %
\begin{figure*}
    \centering
    \includegraphics[width=\textwidth]{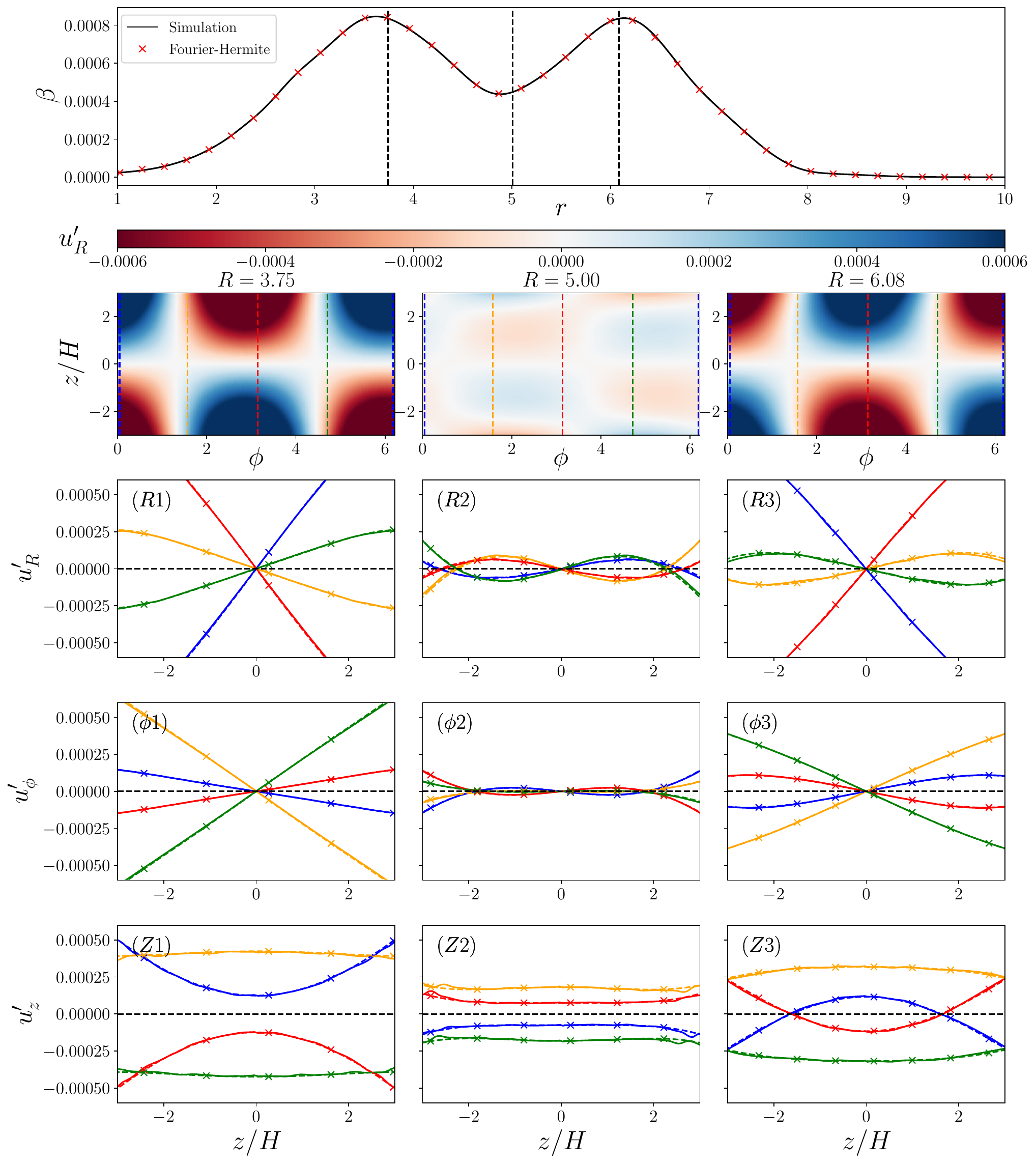}
    \caption{Same as for Fig.~\ref{fig:global_iso_hydro_hermite_panels_0.00175} but for the locally isothermal disc. Note that we don't consider the point mass potential experiment in this case.}
    \label{fig:locally_iso_hydro_hermite_panels_0.00175}
\end{figure*}
% -.-.-.-.-.-.-.-.-.-.-.-.-.-.-.-.- %

In Fig.~\ref{fig:global_iso_hydro_hermite_panels_0.00175} we remedy this for our globally isothermal run with $\beta_\textrm{amp} = 0.00175$, \texttt{giso\_0.00175}. We focus our attention on time $t = 68.7\Omega_0^{-1}$ when the bending wave has partially decomposed into distinct pulses which have propagated away from $r_\textrm{mid}$. The $\beta$ profile at this time is shown in the upper panel where the solid black line denotes the simulation result and the red markers again exemplify the fantastic agreement using the linear Fourier-Hermite theory with $m=1$ and $n = [1,3,5]$. Furthermore, the dotted blue line plots the result from a complementary simulation run which replaces the $\Phi_\textrm{td}$ potential with the full point mass potential $\Phi_\textrm{K}$ and accordingly adjusted equilibrium state. This only leads to subtle differences which validate the use of the thin disc potential throughout this study.

This tilt profile is intersected by three dashed, black vertical lines at cylindrical radii $R = [3.75,5.0,6.08]$, where underneath we perform slices through the velocity perturbations in the respective left, middle and right hand columns. The row of heat maps shows azimuthal-vertical slices of $u_R^\prime$ along cylinders of our \texttt{giso} run, extending up to 3 local scale heights. Each panel clearly shows the development of anti-symmetric sloshing flows which increase in strength above the midplane. Whilst the flow seems to be azimuthally reflected about $\phi = \pi$ for $R = 3.75$, the middle and right hand panels suggest a more deformed structure where the phase of maximal $u_R^\prime$ varies with $z$. This is qualitatively similar to the inclined sloshing mode structures also seen in Fig.~25 of \cite{KimmigDullemond_2024}. To be more precise, we perform slices at four azimuthal quadrants $\phi = [0,\pi/2,\pi,3\pi/2]$ which are marked as dashed blue, yellow, red and green lines respectively. Then in the subsequent three rows we plot the velocity perturbations, $u_R^\prime$, $u_\phi^\prime$ and $u_z^\prime$ at each radial location and similarly coloured azimuthal slice. Each panel is labelled accordingly by the velocity component $(R,\phi,z)$ and the radial slice $(1,2,3)$. Finally, the solid, dashed-crossed and dotted lines show the simulation, Fourier-Hermite theory and point mass potential experiment results respectively. 

Immediately we see remarkable agreement between the Fourier-Hermite theory and our fiducial simulation, with the dashed-crossed lines almost perfectly overlying the solid curves in all panels. At radial positions (1) and (3) we see clear evidence for the linearly increasing sloshing flows proportional to the $z$ coordinate, in the horizontal velocity perturbations $(R,Z)$. These are most prominent as the red and blue curves in (R1) and (R3) at $\phi = [0,\pi]$. Meanwhile the $n=1$ mode seems dominant in the yellow and green lines of ($\phi$1) and ($\phi$3) at $\phi = [\pi/2,3\pi/2]$. This $\pi$ phase shift between the radial and azimuthal perturbations is indicative of the epicylic motions induced by the bending wave \citep{LubowOgilvie_2000}. 

The assumption that $u_R^\prime$ and $u_\phi^\prime$ are $\propto z$ whilst $u_z^\prime$ is vertically constant underpins the reduced bending wave formalism of Section \ref{subsec:reduced_bending}. However, we see clear evidence that higher order polynomial modes are essential in capturing the detailed velocity structure in many of the panels. For example, in (Z1), (Z2) and (Z3) the approximately quadratic profiles of the red and blue lines correspond to compressive vertical `breathing' motions which squash and expand the disc at anti-posed azimuthal phases. These warped breathing motions have been appreciated by \cite{OgilvieLatter2013a, FairbairnOgilvie2021a} and perhaps have profound implications for nonlinear warp amplitudes where the disc becomes highly compressed in a `bouncing' regime \citep{FairbairnOgilvie2021b,HeldOgilvie_2024}. Furthermore in panels (R2) and ($\phi$2) where there is a trough in the tilt profile, we see the velocity perturbations are dominated by the $n=3$ cubic mode. Therefore whilst the structure is locally linear near the midplane, the higher order polynomials can reverse the perturbation sign and lead to a steepening of the curves above $|z| > H$. Whilst the reduced mode formalism can therefore adequately model the velocity structure within one scale height, more significant departures emerge beyond this and demand the generalised Fourier-Hermite method. Nonetheless, since the mass budget is dominated by the dense midplane, the $\beta$ and $\gamma$ diagnostics are heavily weighted by this region and hence sample the linear flow field approximation more faithfully --- hence explaining the success of the reduced bending wave theory despite the presence of higher order modes. In these panels we also note that the dotted curves, corresponding to the use of the point mass potential, also agree rather well with the thin disc simulation and theory, which again verifies the use of the thin disc approximation. The small discrepancies at high values of $z$ for the horizontal velocity components, and the offsets seen in $u_z^\prime$, originate from the slightly different hydrostatic balance incurred between the thin disc and Keplerian potentials. 

We can perform the same analysis for the locally isothermal case which is shown in Fig.~\ref{fig:locally_iso_hydro_hermite_panels_0.00175} (albeit we don't consider a point mass potential run in this case). A similar commentary can then be applied as for the globally isothermal study. Namely, the $n=1$ modes are dominant at the tilt peaks but higher order polynomials are also important, especially at $R=5.0$ for the horizontal velocity perturbations. The locally isothermal Fourier-Hermite theory effectively matches the simulation with a stunning degree of accuracy. Although not plotted in this figure, we have experimented with which locally isothermal terms in equations \eqref{eq:fourier_hermite_uR}-\eqref{eq:fourier_hermite_W} should be included for good agreement. As discussed in Section \ref{subsec:reduced_bending} the $s$ term in \eqref{eq:fourier_hermite_uR} and the $s h$ term in \eqref{eq:fourier_hermite_uphi} are crucial for capturing the qualitative bending wave evolution. Retaining these but dropping the other $h^2$ terms gives decent agreement between the velocity perturbation profiles within $1H$. However, the neglected $h^2$ terms are necessary for a good match at higher latitudes. Once again, these details for the velocity profile in the lower density disc atmospheres, are suppressed by the integrated $\beta$ diagnostic and hence have not been appreciated in previous studies.

% ........................................ %
\subsection{Weak nonlinearities}
\label{subsec:weak_nonlinearities}
% ........................................ %

% -.-.-.-.-.-.-.-.-.-.-.-.-.-.-.-.- %
\begin{figure*}
    \centering
    \includegraphics[width=\textwidth]{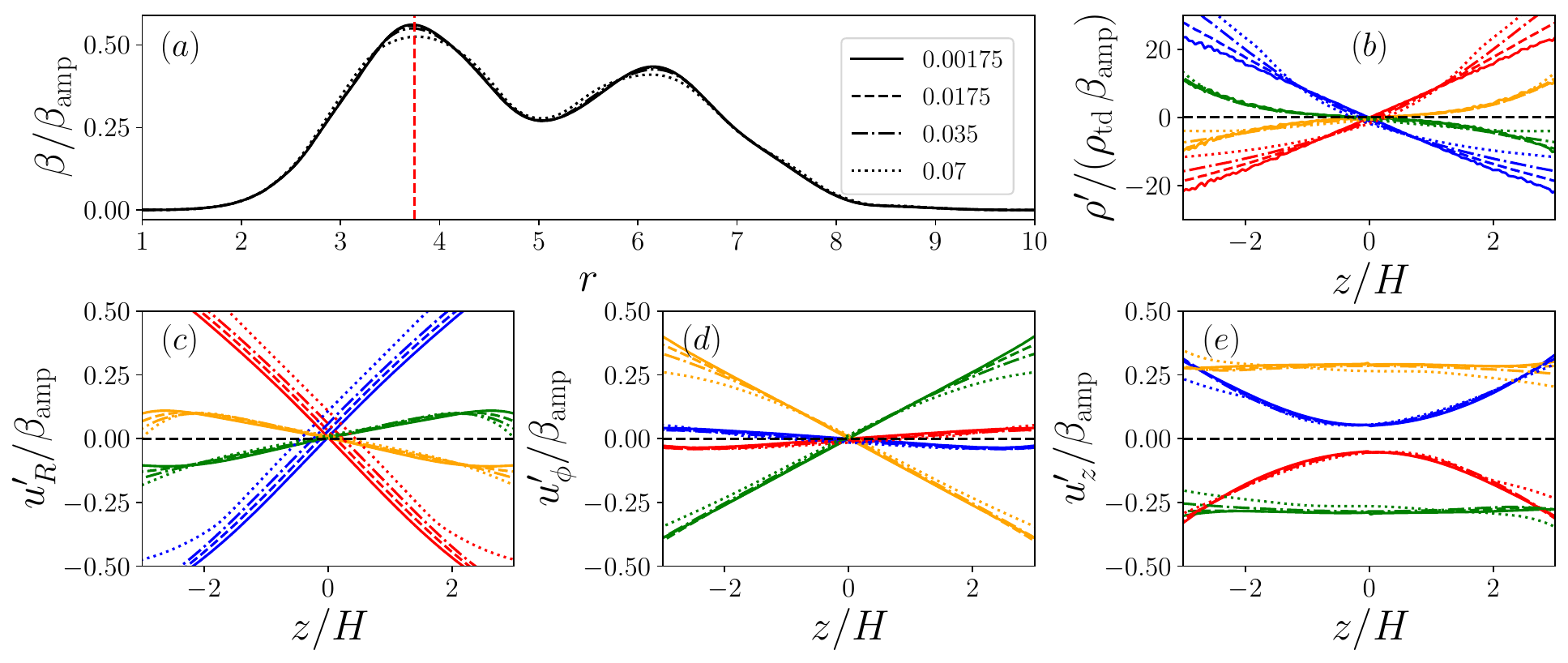}
    \label{fig:giso_nonlinearity}
    \caption{Variation of the normalised perturbed profiles for $\beta_\textrm{amp} = [0.00175,0.0175,0.035,0.07]$. The corresponding line styles are denoted in the inset legend. (a) normalised tilt profiles at $t = 68.7\Omega_0^{-1}$. The dashed red line denotes $R = 3.75$ where we examine the scaled perturbations in (b) $\rho^\prime$ (c) $u_R^\prime$ (d) $u_\phi^\prime$ and (e) $u_z^\prime$. These vertical profiles are plotted for 4 azimuthal quadrants $\phi = [0,\pi/2,\pi/3\pi/2]$ shown in blue, orange, red and green respectively.}
\end{figure*}
% -.-.-.-.-.-.-.-.-.-.-.-.-.-.-.-.- %

So far we have focused on the fiducial case study with $\beta_\textrm{amp} = 0.00175$ which resides confidently within the linear perturbative regime. Now we will consider our globally isothermal simulation results as we boost $\beta_\textrm{amp}$ towards weakly nonlinear tilt amplitudes, for which $\beta \sim H/R$. In Fig.~\ref{fig:giso_nonlinearity} we examine the warp induced perturbation profiles for the \texttt{giso} runs which are normalised by the tilt amplitude. Panel (a) plots the normalised tilt profile $\beta/\beta_\textrm{amp}$ at $t = 68.7\Omega_0^{-1}$ for 4 different values of $\beta_\textrm{amp} = [0.00175,0.0175,0.035,0.07]\,\textrm{radians} = [0.1,1.0,2.0,4.0]\,\textrm{degrees}$. The corresponding line-styles are marked in the accompanying legend. The tilt profile shape remains largely the same, indicating that the linear approximation is working well even up to $4^\circ$. At this highest value of the tilt amplitude there are small deviations in the profile but it still agrees remarkably well. 

To gain further insight, one can perform vertical slices through perturbed quantities at the cylindrical radius $R = 3.75$ (marked by the dashed red line in panel (a)) and $\phi = [0,\pi/2,\pi,3\pi/2]$ (blue, orange, red and green lines). Panel (b) plots the relative density perturbation normalised by the tilt amplitude whilst panels (c), (d) and (e) plot the normalised $R$, $\phi$ and $z$ velocity perturbations respectively. As the tilt is increased, the curves mostly overlap within $1H$ although differences do start to develop at higher latitudes. Furthermore, there seems to be a widening offset between line-styles for the blue and red curves in panel (c). This owes to the growth of an $n=0$, $m=2$ mode which is not computed for in our purely linear theory which considered only $n=1$ and $m = [1,3,5]$. Nonetheless, these even-parity vertical modes cancel out their integrated contribution to the horizontal angular momentum components and therefore don't lead to notable differences in the $\beta$ profile.

One might be surprised that the profiles scale so well according to linear theory. Indeed, scaling out the normalisation factor from panel (b), we can see that even for $\beta_\textrm{amp} = 0.0175$ the relative density perturbation begins to approach order unity for larger values of $z$. As noted by \cite{LubowOgilvie_2000}, the linear predictions outperform their formal range of validity within the Eulerian analysis, which technically requires $\beta \ll H/R$. Indeed, the steady state linear rigid tilt solution can of course be extended towards arbitrary amplitude in a spherically symmetric potential even though the Eulerian perturbations become large. An alternative Lagrangian perspective might be more suitable here, wherein perturbations are considered relative to material elements in the underlying flat disc \citep{FriedmanSchutz_1978}.

Another related subtlety should also be acknowledged. In our linear analysis, we project the exact initial conditions described by equations \eqref{eq:ic_rotate_rho}-\eqref{eq:ic_rotate_u}, onto the Fourier-Hermite basis and retain only the $m = 1$ and $n=[1,3,5]$ modes. In reality, the initial condition will also have subdominant projections onto other neglected modes. It can be shown that whilst the odd $n$ perturbations scale linearly with $\beta_\textrm{amp}$, the even $n$ modes scale as $\beta_\textrm{amp}^2$. Therefore, these even contributions can safely be ignored when we are firmly in the linear regime. However, as the tilt increases one might be tempted to include them. However, doing so yields much poorer agreement compared with the simulations which are still dominated by the odd vertical modes as seen in Fig.~\ref{fig:giso_nonlinearity}. In this sense, the nonlinearity projected from the initial condition is effectively counterbalanced by the nonlinearity in the full numerical equations. Meanwhile, including the $\mathcal{O}(\beta_\textrm{amp}^2)$ modal projections is not entirely self-consistent within the linearized framework --- the linear theory is unable to temper their growth and this leads them to becoming overly pronounced in the evolved perturbation.
% ============================ %
\section{MRI Turbulent Method} 
\label{sec:mri_method}
% ============================ %

Having examined bending wave propagation in hydrodynamical, laminar discs we are now interested to discover how the picture changes when these disturbances are launched in turbulent flows. To generate this turbulence we will appeal to the \textit{magnetorotational instabilty} (MRI) which has long been lauded as a means of driving accretion \citep{BalbusHawley_1991,BalbusHawley_1992,HawleyBalbus_1991,HawleyBalbus_1992}. To this end we will describe the magnetohydrodynamic (MHD) equations being solved and the numerical setup in Section \ref{subsec:mhd_setup}, before describing how the ensuing turbulent state is mapped towards a controlled bending wave initial condition in Section \ref{subsec:MHD_bending_launch}.

% ---------------------------- %
\subsection{Numerical implementation of the MRI}
\label{subsec:mhd_setup}
% ---------------------------- %

\texttt{Athena++} is built with MHD in mind and has been a powerful code for investigating magnetised turbulence. The previous hydrodynamic equations \eqref{eqn:athena_continuity}-\eqref{eqn:athena_eos} are now generalised such that the momentum equation becomes 
\begin{equation}
    \label{eqn:athena_mhd_momentum}
    \frac{\partial (\rho \mathbf{v})}{\partial t} + \nabla\cdot\left[ \rho \mathbf{v}\mathbf{v}-\mathbf{B}\mathbf{B}+P^{*}\mathbf{I}\right]+\rho\nabla\Phi_\textrm{td} = 0,
\end{equation}
where $P^{*} = P+\frac{\mathbf{B}\cdot\mathbf{B}}{2}$ and $\mathbf{B}$ is the magnetic field. The magnetic field itself is updated according to the induction equation
\begin{equation}
    \label{eqn:athena_mhd_induction}
    \frac{\partial B}{\partial t} - \nabla\times\left[\mathbf{v}\times\mathbf{B}-\eta_\textrm{O}(\nabla\times\mathbf{B}) \right]= 0,
\end{equation}
where $\eta_\textrm{O}$ is the Ohmic diffusivity. 

Our numerical setup is similar to the hydrodynamical case study described in Section \ref{sec:hydro_method} with the magnetic extensions inspired largely by the global MRI configurations of \cite{FromangNelson_2006} and \cite{FlockEtAl_2011}. The hydrodynamic variables are set according to the flat disc equilibrium, within the same physical domain range considered previously. We adopt identical hydro parameters and consider a globally isothermal run with $d = 2$, $s = 0$ and $c_{0} = 0.035$. Now we also implement an initially weak toroidal magnetic field with plasma beta $\beta_\textrm{m} \equiv 2P/B^2 = 25$. This is contained within the meridional region bounded radially by $r = [2.1,8.9]$ and vertically by $\pm 3H$. In order to limit the interaction with the boundaries we design radial buffer zones for $r < 2.0$ and $r > 9.0$ which act to damp out disturbances. Specifically, we relax the density and momentum back towards the flat disc equilibrium according to the prescription
\begin{equation}
    \label{eq:rho_relax}
    \frac{d\rho}{dt}\bigg|_\textrm{relax} = -\frac{\Omega_\textrm{K}}{\tau}(\rho-\rho_\textrm{td}) ,
\end{equation}
and similarly for the momentum. Here, $\tau$ sets the relaxation timescale in local dynamical units and is chosen to be $1$. Meanwhile we also set $\eta_\textrm{O}$ to be a linearly increasing function from $0.0$ at the interior buffer interfaces to $4.9\times 10^{-4}$ at the disc edge, diffusing $\mathbf{B}$ which might otherwise become wound up by the boundaries. Furthermore, throughout the physical domain we also enact an anomalous diffusivity $\eta_\textrm{O} = 4.9\times 10^{-4}$ which is switched on whenever $\beta_\textrm{m} < 10^{-3}$. This helps suppress the Alfvén velocity in the low density disc atmosphere, which otherwise acts as a severe limitation on the Courant-Friedrichs-Lewy (CFL) timestep. Additionally, between the buffer zones we also implement a slower density relaxation source term according to equation \eqref{eq:rho_relax} with a much longer characteristic timescale, $\tau = 10$. This acts to maintain the initial surface density profile and counteract the effects of accretion, which otherwise would significantly modify the disc background upon which we want to launch bending waves. Finally, in order to seed the growth of the MRI, we add random white noise to the meridional velocity components, sampled uniformly between $\pm 0.025 c_{0}$. Once the MRI takes hold, the details of this perturbation are forgotten.

Our boundaries are slightly modified in the polar direction, which are now set to diode conditions. This enforces all hydro quantities to be zero-gradient across the boundary apart from the normal component of velocity which permits a zero-gradient outflow whilst preventing inflows via a reflective condition. Meanwhile for $\mathbf{B}$, the radial and polar boundaries extrapolate the tangential components of the magnetic field into the ghost zones according to the force free profile $\propto 1/r$. The remaining normal component is then fixed by the solenoidal constraint, $\nabla\cdot\mathbf{B} = 0$.

For these runs we will adopt a uniform grid with $(N_r,N_\theta,N_\phi) = (384,216,768)$ which is comparable to the highest resolution, global run performed by \cite{FlockEtAl_2011}. They comment that a sufficiently high resolution of $\sim 25$ cells per $H$ is desirable to attain a self-sustaining zero-net flux MRI. We achieve this criterion at $r_\textrm{mid} = 5$ and our results in Section \ref{subsec:MRI_charactersation} suggest a well-behaved MRI is obtained. During our testing we have examined lower resolution runs which tend to saturate in states with reduced levels of turbulence, in keeping with the findings of \cite{FlockEtAl_2011}. Whilst we will characterise the turbulent state in some detail, the focus of our investigation is not about obtaining a highly converged MRI, but rather developing a turbulent background which will interact with bending waves.

% ---------------------------- %
\subsection{Launching bending waves}
\label{subsec:MHD_bending_launch}
% ---------------------------- %

After obtaining a developed state of turbulence, we are in a position to launch bending waves. We will now outline the recipe for doing this. Firstly, the turbulence itself leads to some inherent noise in the $\beta$ and $\gamma$ profiles, even for the initially flat disc. We rigidly rotate the density and velocity variables in spherical shells to flatten this out so that $\beta(r) = 0$. Note that during this flattening process we do not change the twist in order to avoid shearing out the perturbations azimuthally. Afterwards, we compose this with another radially dependent rotation, which maps the density and velocity to the target bending wave state. This is essentially the same process as described in Section \ref{subsec:hydro_ics} equations \eqref{eq:ic_rotate_rho}-\eqref{eq:ic_rotate_u}, which transforms the disc to the the tilt profile described by equation \eqref{eq:target_profile}.

A slight complication is how to treat the magnetic field. If we were to apply the same rotation to $\mathbf{B}$ as for the velocity field $\mathbf{u}$, we would unwittingly break the solenoidal constraint when this process is discretized over our finite volume grid. Since, we are only considering very small warps in this paper, we will instead elect to leave the magnetic field untouched. This is valid since over the warping scale considered, the structure of the magnetised turbulence doesn't change significantly in character. Should we wish to consider larger warp amplitudes in the future, one must employ a more careful method which rotates the magnetic field. This might be enacted through a post-rotation  `cleaning' of the non-zero divergence. Alternatively one could impose an artificial velocity field in the code which advects the disc to the desired state whilst the concurrent numerical solution of the induction equation preserves the solenoidal constraint. Such endeavours will be left to future investigations.

With this initial condition, the other numerical considerations remain essentially the same when we restart the run. However, we now switch off the slow density relaxation source term throughout the physical domain, which would otherwise alter the angular momentum transport we wish to probe.

% ============================ %
\section{MRI turbulent bending waves} 
\label{sec:mri}
% ============================ %

In this section we will examine the profound effect that MRI turbulence has on the propagation of bending waves. In Section \ref{subsec:MRI_charactersation} we will characterise the development of the turbulence, the scales it operates on and quantify the feedback onto the disc in terms of correlated stresses. Then in Section \ref{subsec:turbulent_bending} we will examine how bending waves of different amplitude interact with the turbulence, comparing it with the laminar evolution studied extensively in Section \ref{sec:laminar_results}. 

% ---------------------------- %
\subsection{Characterising the MRI}
\label{subsec:MRI_charactersation}
% ---------------------------- %

% -.-.-.-.-.-.-.-.-.-.-.-.-.-.-.-.- %
\begin{figure*}
    \centering
    \includegraphics[width=\textwidth]{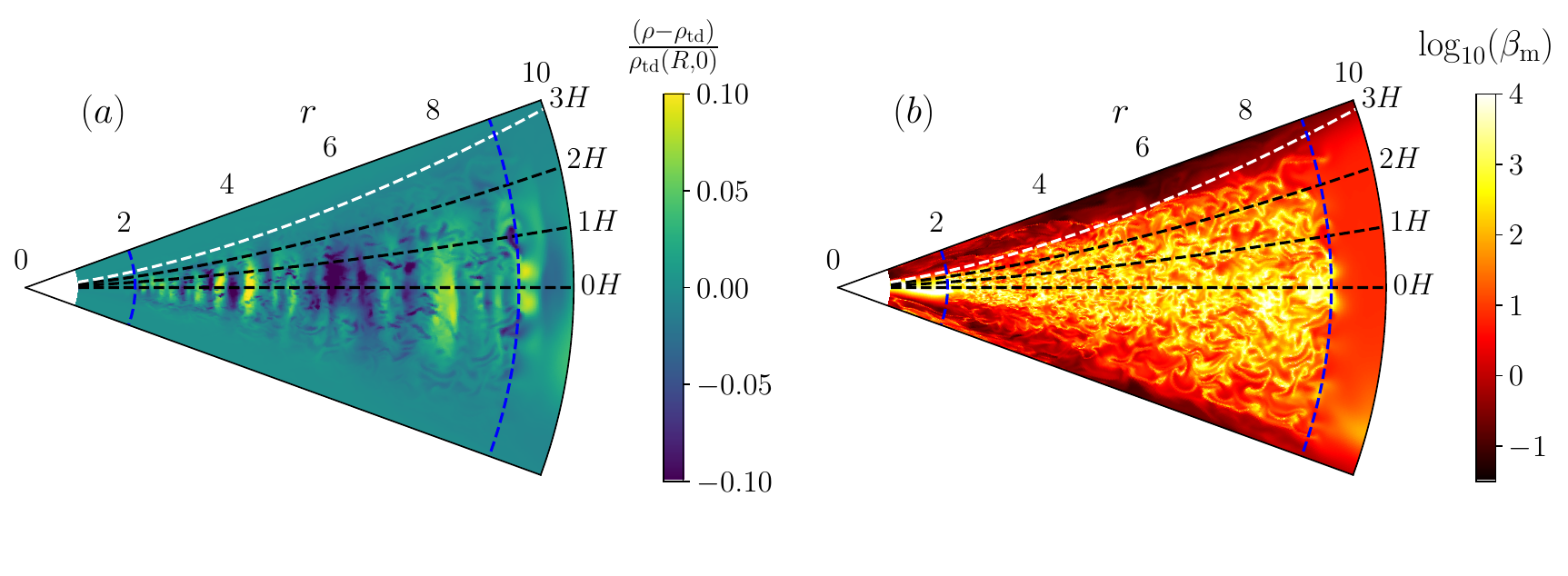}
    \caption{Turbulent structure in meridional slice through $\phi = 0$ at $t = 2953\Omega_0^{-1}$. Panel (a) shows $(\rho-\rho_\textrm{td})/\rho_{\rm td}(R,0)$. Panel (b) shows $\log_{10}(\beta_\textrm{m})$. The dashed black and white curves mark the latitude measured in integer steps of $H$. The blue dashed lines mark the inner and outer buffer zone edges.}
    \label{fig:turbulent_snapshot}
\end{figure*}
% -.-.-.-.-.-.-.-.-.-.-.-.-.-.-.-.- %

The setup described in Section \ref{subsec:mhd_setup} is run for 635 inner orbital periods $T_0$, or equivalently 20 outer orbital periods. This gives sufficient time for the MRI to grow locally over several dynamical timescales and then saturate globally. A meridional slice through $\phi = 0$ is visualized in Fig.~\ref{fig:turbulent_snapshot} for $t = 2953\Omega_0^{-1} = 470 T_0$. 

In panel (a) we plot the density perturbations normalized by the equilibrium midplane density, $(\rho-\rho_\textrm{td})/\rho_{\rm{td}}(R,0)$, which is chosen to highlight the formation of turbulent structure. This reveals vertical bands of over/under densities, typical of radially propagating density waves which are also found in Fig.~4 of \cite{FromangNelson_2006} for example. The buffer zones are demarcated by blue dashed lines where the density relaxation acts to damp these disturbances. Meanwhile in panel (b) we plot $\log_{10}(\beta_\textrm{m})$ which clearly demonstrates the finely structured turbulence as the magnetic field is twisted up by the velocity perturbations. The disc remains pressure dominated near the midplane where $\beta_\textrm{m}$ tends to be $> 10$, but it can drop below unity in the corona for $|z/H|>3$. 

In order to quantify the strength of the turbulence we will now compute an effective $\alpha$ stress parameter \citep[c.f.][]{BalbusPapaloizou_1999,Balbus_2003}. We start by defining the density weighted azimuthal-vertical average as 
\begin{equation}
    \label{eq:density_shell_avg}
    \langle X \rangle_\rho = \frac{\iint \rho X r^2 \sin\theta \, d\theta \, d\phi}{\iint \rho r^2 \sin\theta \, d\theta \, d\phi} ,
\end{equation}
and the density weighted volume average as 
\begin{equation}
    \label{eq:density_vol_avg}
    \langle X \rangle_V = \frac{\iint \rho X r^2 \sin\theta \, dr \, d\theta \, d\phi}{\iint \rho r^2 \sin\theta \, dr \, d\theta \, d\phi} .
\end{equation}
We perform these integrals over the full azimuthal domain, the radial range between the buffer zones $r = [2,9]$ and the polar extent $\theta = \pm 3 H(r)/r$, where the scale height defined by equation \eqref{eq:scale_height} is presently evaluated at the numerical value of the local spherical (rather than cylindrical) radius. We then compute the radial stress profiles as
\begin{equation}
    \label{eq:radial_alphas}
    \alpha_\textrm{R}(r) = \frac{\langle u_R^\prime u_\phi^\prime\rangle_\rho}{c^2 q} \quad \textrm{and} \quad \alpha_\textrm{M}(r) = -\frac{\langle B_R B_\phi/\rho \rangle_\rho}{c^2 q} , 
\end{equation}
and the volumetric averages
\begin{equation}
    \label{eq:volume_alphas}
    \alpha_\textrm{R,V} = \frac{\langle u_R^\prime u_\phi^\prime\rangle_V}{c^2 q} \quad \textrm{and} \quad \alpha_\textrm{M,V} = -\frac{\langle B_R B_\phi/\rho \rangle_V}{c^2 q}.
\end{equation}
Here, we have extracted the perturbed velocity via $u_R^\prime = v_R-\langle v_R\rangle_\rho$ and $u_\phi^\prime = v_\phi-\langle v_\phi\rangle_\rho$. Furthermore, $q = -d \ln\Omega_\textrm{K}/ d\ln R = 3/2$ defines the Keplerian orbital shear parameter. Note that the simulation vector quantities are first converted into cylindrical components whilst the weighted integrals are still performed over the numerical spherical grid\footnote{For such a thin disc, integrating over cylinders would give a very similar result but complicates the analysis since interpolation onto a new grid is required}. The subscript `R' denotes the Reynolds stresses whilst the subscript `M' refers to Maxwell stresses. The net stress measure is then given by $\alpha = \alpha_\textrm{R}+\alpha_{M}$. These definitions are motivated by equating the turbulent correlations in the $R$-$\Phi$ direction to an effective Shakura-Sunayaev viscous stress. Here we only consider the standard, horizontal stress correlations which are associated with driving accretion. Of course, the evolution of the warp will also engage mixed vertical-horizontal stresses which may be different in character. Indeed, several studies suggest that MRI turbulence is highly anisotropic with power concentrated along the $k_z$ direction \citep[e.g.][]{HawleyEtAl_1995,MurphyPessah_2015}. Additionally, in the local shearing box models of unstratified MRI turbulence, one might not expect any averaged vertical stresses since there is no preferential direction set by the background equilibrium state. However, \cite{TorkelssonEtAl_2000} find that when a sloshing motion is inserted into a stratified, MRI turbulent shearing box, the epicyclic motions damp on a time consistent with adopting an isotropic viscous stress, as measured from the $R$-$\phi$ stress component. 

% -.-.-.-.-.-.-.-.-.-.-.-.-.-.-.-.- %
\begin{figure}
    \centering
    \includegraphics[width=\columnwidth]{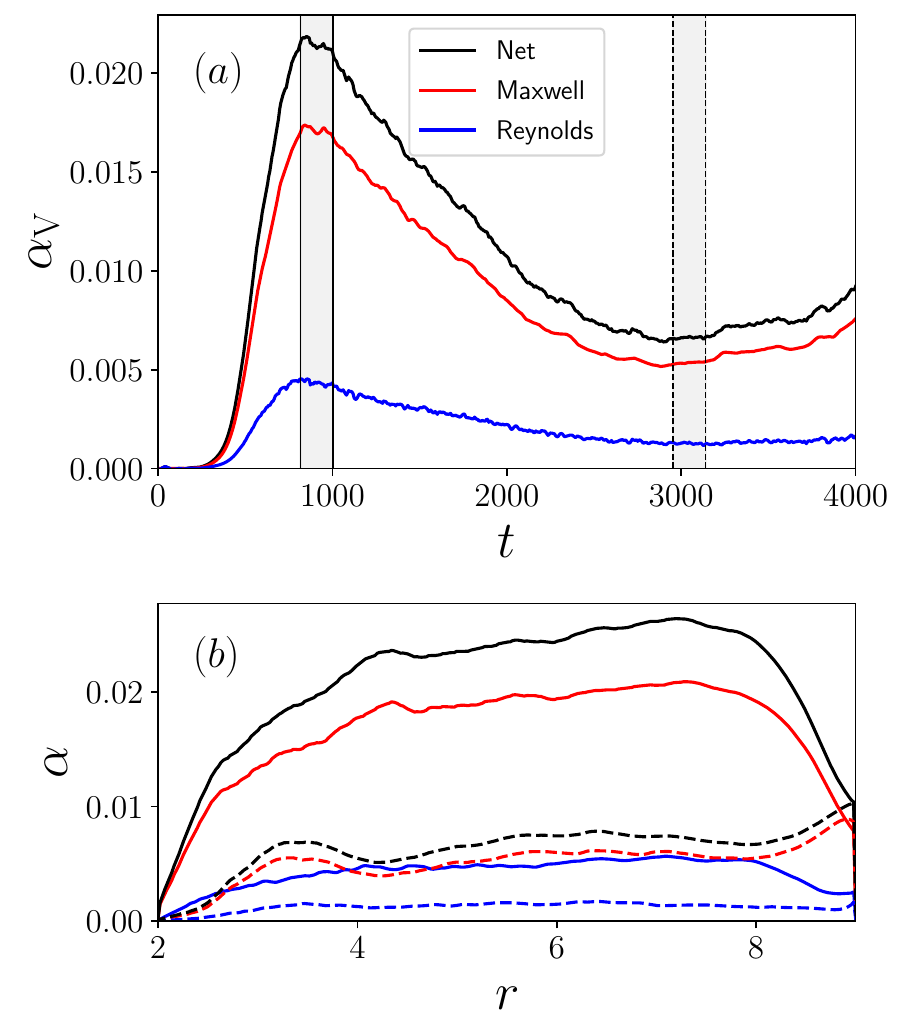}
    \caption{Panel (a): net volume averaged stress profile (black) decomposed into the Maxwell (red) and Reynolds (blue) contributions. High and low turbulent intervals are marked between $[817,1005]\Omega_0^{-1}$ and $[2953,3142]\Omega_0^{-1}$. Panel (b): corresponding radial stress profile decomposition averaged over these high (solid lines) and low states (dashed lines).}
    \label{fig:stress_profile}
\end{figure}
% -.-.-.-.-.-.-.-.-.-.-.-.-.-.-.-.- %

In Fig.~\ref{fig:stress_profile} we examine the evolution of $\alpha$ for our simulation. In panel (a) we plot the average volumetric stress parameters, with the black net $\alpha_V$ line decomposed into the red $\alpha_{M,V}$ and blue $\alpha_{R,V}$ lines. Consistent with previous studies, we observe an initially steep rise in $\alpha_V$ as the linear MRI grows throughout the disc. By around $t = 1000\Omega_0^{-1} = 160 T_0$ the linear instability has had time to saturate in the outer disc, the turbulence then peaks and begins to turnover as magnetic flux leaves the system. This decay eventually levels out around $t = 3000 \Omega_0^{-1} = 480 T_0$ where the turbulence seems to inhabit a quasi-steady state. Clearly across all times, $\alpha_V$ is dominated by the Maxwell component which is roughly 4 times greater than the Reynolds contribution after the initial turbulent peak. Towards, the end of the simulation there is evidence of another rise phase as the MRI begins to grow again. This is suggestive of longer term variability which may in part be triggered by our slow density relaxation source term -- bringing the disc back towards the initial condition which is susceptible to another phase of increased instability. Of course, such high and low turbulent accretion states are an inherent part of modelling a number of variable sources. We will continue to refer to the \textit{High State} as the shaded time interval bounded by the solid vertical lines between 817 and 1005 $\Omega_0^{-1}$. Meanwhile the \textit{Low State} is marked between the dashed lines at 2953 and 3142 $\Omega_0^{-1}$. The time averaged value of the net volumetric stress during this high state is $\alpha_{V,H} = 0.022$, which is 3 times greater than the low state value, $\alpha_{V,L} = 0.0066$.

In panel (b) we examine the radial stress profiles averaged over these high and low time intervals, indicated by the solid and dashed lines respectively. We see that throughout the radial domain, the Maxwell stress is $\sim 4$ times the Reynolds stress in keeping with the volumetric average results. These profiles exhibit an approximately constant plateau within the bulk of the physical domain which perhaps validates the oft used spatially uniform $\alpha$ approximation. However, we do acknowledge the drop off near the inner and outer edges where the turbulence feels the effects of the buffer zone damping.

% -.-.-.-.-.-.-.-.-.-.-.-.-.-.-.-.- %
\begin{figure*}
    \centering
    \includegraphics[width=\textwidth]{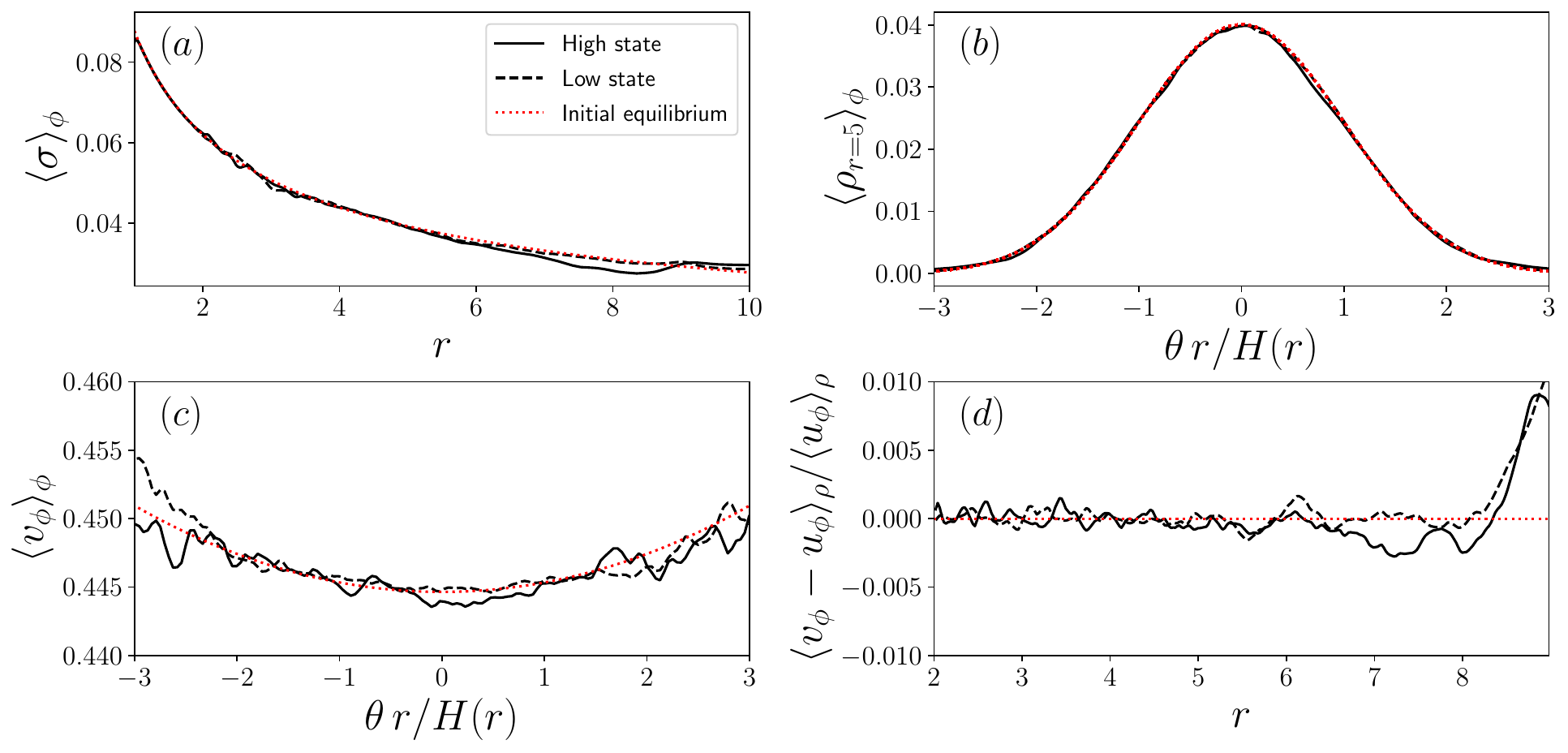}
    \caption{Comparison of turbulent disc profiles during the high state snapshot ($t = 817\Omega_0^{-1}$; solid line) and low state snapshot ($t = 2953\Omega_0^{-1}$; dashed line), with the disc equilibrium (dotted red line). Panel (a): azimuthally-averaged surface density. Panel (b): azimuthally-averaged vertical density profile at $r = 5$. Panel (c): azimuthally-averaged vertical $v_\phi$ profile at $r=5$. Panel (d): mass weighted, shell-averaged $v_\phi$ perturbation relative to the equivalent average over the equilibrium disc.}
    \label{fig:turbulent_background}
\end{figure*}
% -.-.-.-.-.-.-.-.-.-.-.-.-.-.-.-.- %

It is also worthwhile checking whether there is any significant evolution of the background disc profile during these high and low states. Indeed, we would like the mean structure to be broadly preserved for better comparison with the laminar bending waves investigated in the previous sections. In Fig.~\ref{fig:turbulent_background} all panels show solid and dashed lines corresponding to disc profiles captured during the high state ($t = 817\Omega_0^{-1}$) and low state ($t = 2953\Omega_0^{-1}$) respectively. Meanwhile the dotted red line denotes the initial equilibrium profile. In addition to the averaging operators introduced via equations \eqref{eq:density_shell_avg}-\eqref{eq:density_vol_avg}, we will also make use of the simple azimuthal average $\langle \cdot \rangle_\phi = \frac{1}{2\pi}\int_{0}^{2\pi} \cdot \, d\phi$.

In panel (a) we see that the equilibrium azimuthally averaged surface density curve is closely traced by the turbulent profiles. In particular, the low state seems to closely overlie the target density as the slow density relaxation process effectively counteracts the turbulent accretion. In panel (b) we also plot the azimuthally averaged vertical density at $r = 5$ across 3 scale heights. Clearly the Gaussian hydrostatic profile is well preserved with only slight deviations discernible in the high turbulence state. More prominent perturbations can be seen in the azimuthal velocity profile at $r = 5$, as detailed in panel (c). Recall that for a globally isothermal disc the equilibrium $v_\phi$ is constant on cylinders. However, since the spherical coordinate system intersects different $R$ as a function of $\theta$ for fixed $r$, we see the slight curve in the equilibrium profile. The azimuthally averaged $v_\phi$ closely matches this equilibrium near the midplane, with larger perturbations towards the low density regions at a few scale heights. Finally, panel (d) presents an alternative measure of the azimuthal perturbations --- the normalised discrepancy between the density weighted shell average azimuthal perturbation, $\langle v_\phi-u_{\phi} \rangle_\rho / \langle u_{\phi} \rangle_\rho$. The averaged perturbations are $\ll 1\%$ of the near-Keplerian background throughout the disc, with the smallest deviations at low $r$ growing towards the outer buffer zone interface. Although these relative perturbations may seem small, since the tilt amplitudes for linear bending waves considered in Section \ref{sec:laminar_results} are $\lesssim H/R$, such disturbances might still significantly modify the warp evolution. 

% -.-.-.-.-.-.-.-.-.-.-.-.-.-.-.-.- %
\begin{figure}
    \centering
    \includegraphics[width=\columnwidth]{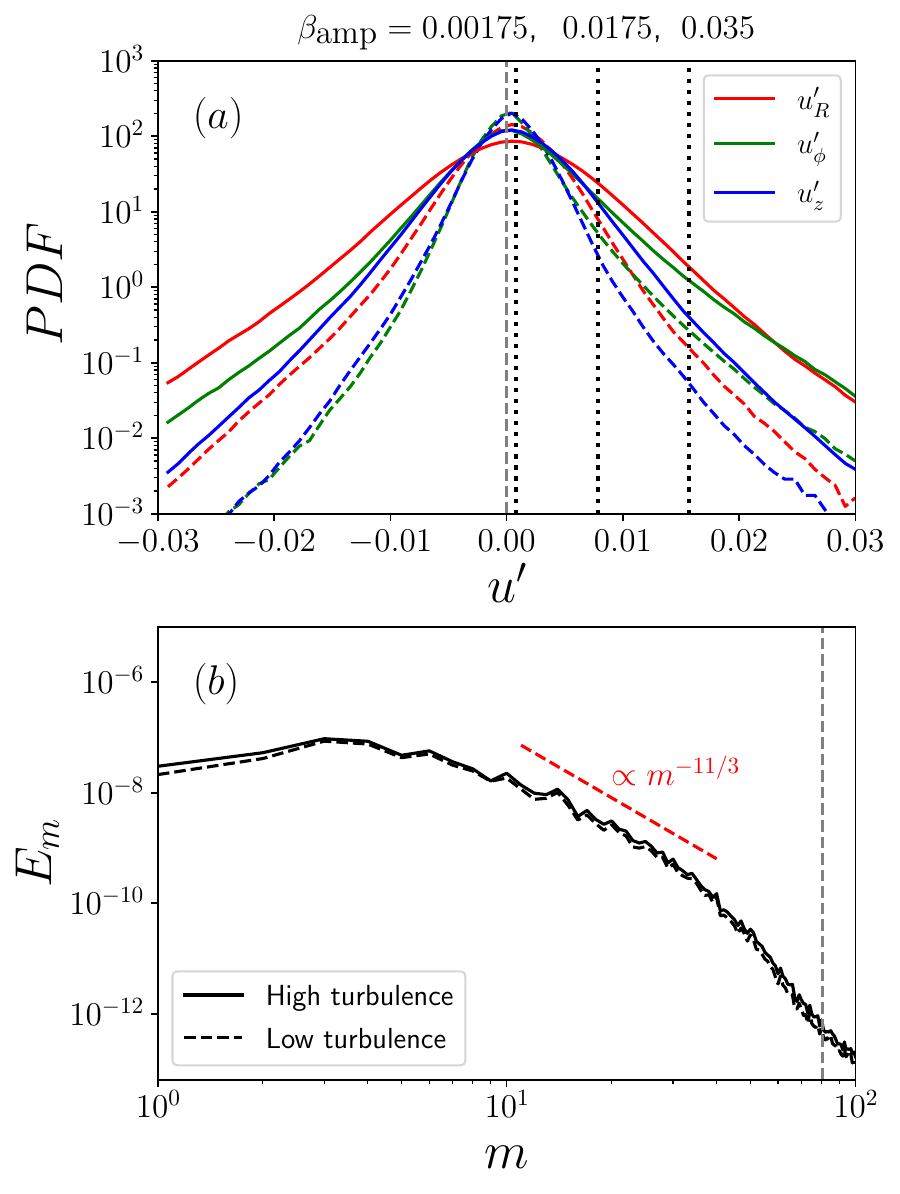}
    \caption{Velocity perturbation analysis for the high (solid) and low (dashed) states. \textit{(a)} mass weighted probability density function for velocity perturbations $u_R^\prime$ (red), $u_\phi^\prime$ (green) and $u_z^\prime$ (blue). The vertical dotted lines denote characteristic bending wave velocities associated with the tilt amplitudes marked above. \textit{(b)} azimuthal slice through perturbation power spectrum, averaged over the high and low turbulent intervals. The Kolmogorov slope is indicated as a dashed red line whilst the $H(r_{\rm mid}=5)$ characteristic length scale is shown as the grey vertical line.}
    \label{fig:turbulent_power}
\end{figure}
% -.-.-.-.-.-.-.-.-.-.-.-.-.-.-.-.- %

To determine the characteristic magnitude and length scales of the turbulent velocity field in more detail we turn to Fig.~\ref{fig:turbulent_power}. In panel (a) we have computed the mass weighted probability density function for the cylindrical velocity perturbations $u^\prime = v - \langle v \rangle_\rho$. The $R$, $\phi$ and $z$ components are shown in red, green and blue respectively. The solid lines are extracted for the $t = 817\Omega_0^{-1}$ high state snapshot whilst the dashed lines correspond to the low state $t = 2953\Omega_0^{-1}$ snapshot. All distributions peak around the zero equilibrium and are approximately symmetric (although a slight skew towards negative radial velocities is indicative of the turbulent accretion flow). In both states, the horizontal perturbations exhibit a wider spread, once again hinting at the fundamental anisotropy of the underlying turbulence. Furthermore, the low state presents a significantly narrower distribution when compared with the high state, consistent with the reduced levels of turbulence in the disc at this later time. To put this into perspective, we have over-plotted dotted vertical lines which are indicative of the typical vertical perturbation induced by a laminar bending wave, $u_z^\prime \sim r_\textrm{mid}\Omega(r_\textrm{mid}) \beta_\textrm{amp}$ (see Section \ref{subsec:reduced_bending}). Moving from the centre outwards, we plot this characteristic value for $\beta_\textrm{amp} = [0.00175,0.0175,0.035]$. Clearly for the smallest tilt amplitude, the turbulent distribution far exceeds the characteristic warping flows and therefore we expect the bending wave to be rapidly disrupted. Meanwhile, upon reaching the larger tilt velocities, the turbulent PDF has dropped by $\sim 1$ or 2 orders of magnitude and hence we expect the bending wave to be survive for longer. These predictions will be confirmed in Section \ref{subsec:turbulent_bending}. In panel (b) we instead examine the typical spatial scales of the turbulence. The Fourier transform of the cylindrical velocity perturbations is first computed along the azimuthal direction and then averaged over the meridional area $A_\textrm{merid}$, bounded by $r = [3,7]$ and $\theta = \pi/2 \pm 0.2$, yielding
\begin{equation}
    \label{eq:fourier_m}
    \tilde{u}^\prime(m)= \frac{1}{2\pi \,A_\textrm{merid}}\iiint u^\prime e^{-i m \phi} \, d\phi \,dA_\textrm{merid}
\end{equation}
This allows us to define a 1D slice through the full 3D power spectrum as
\begin{equation}
    \label{eq:power_spectrum}
    E_m = |\tilde{u}_R^\prime|^2+|\tilde{u}_\phi^\prime|^2+|\tilde{u}_z^\prime|^2.
\end{equation}
We reduce noise by averaging over the duration of the high and low state time intervals, before plotting the corresponding curves as the solid and dashed black lines. Both spectra nearly overlap indicating that the spatial scale of the turbulence is similar in the high and low states. The curves peak at around $m=4$ suggesting that the turbulence is organised over fairly large length scales. Beyond this, there seems to be evidence for a brief inertial range with a Kolmogorov like $m^{-11/3}$ slope between $m = 10$-$40$ before steepening towards the dissipative scales \citep{HawleyEtAl_1995}. For reference, the dashed grey line marks the value of $m$ which sets azimuthal wavelengths at $r_{\rm mid}=5$ equal to the local value of $H$. The typical azimuthal length scales are clearly $> H$ and hence the assumption of localised angular momentum transport, inherent to $\alpha$ models, is questionable. Furthermore, since turbulent eddy sizes are vertically restricted by the scale height, there is a clear spatial anisotropy which is often neglected in viscous models.

% ---------------------------- %
\subsection{Turbulent Bending Waves}
\label{subsec:turbulent_bending}
% ---------------------------- %

\begin{figure*}
    \centering
    \includegraphics[width=\textwidth]{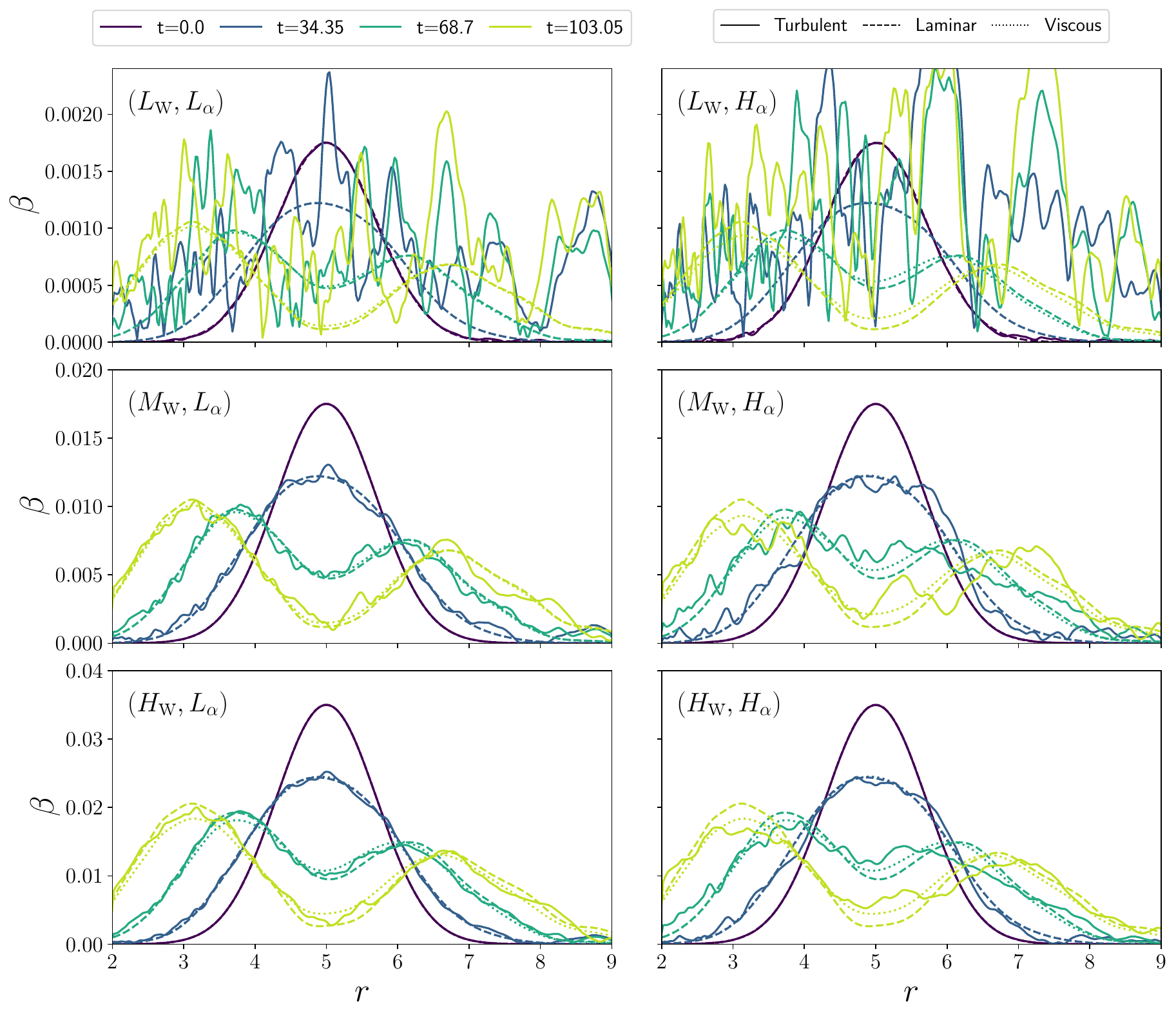}
    \caption{Tilt profiles at 4 different times (see inset coloured legend) for the turbulent (solid), inviscid laminar (dashed) and viscous laminar (dotted) runs. Three tilt initialisations are denoted $(L_\textrm{w}, M_\textrm{w}, H_\textrm{w})$ for $\beta_\textrm{amp} = (0.00175,0.0175,0.035)$. Runs conducted in the low or high turbulent states are denoted by $L_\alpha$ and $H_\alpha$ labels. For the $L_\alpha$ viscous runs we adopt $\alpha = 0.0066$, whilst for $H_\alpha$ we set $\alpha = 0.022$. More details can be found in Section \ref{subsec:turbulent_bending}.}
    \label{fig:turb_bending}
\end{figure*}

Having characterised the base MRI turbulence, we now initialise the disc with bending waves as described in Section \ref{subsec:MHD_bending_launch}. We will consider three tilt amplitudes $\beta_\textrm{amp} = [0.00175,0.0175,0.035]$ launched at the beginning of the high and low states and followed for a time interval of $103.05\Omega_0^{-1}\sim 16 T_0$ until the bending waves begin to interact with the radial buffer zones. The composition of these low ($L_\textrm{w}$), medium ($M_\textrm{w}$) and high ($H_\textrm{w}$) warps with the low ($L_\alpha$) and high ($H_\alpha$) turbulent states yields a space of 6 run configurations which are visualised in Fig.~\ref{fig:turb_bending}. In each panel the different coloured lines denote four equally spaced snapshots between $t=0.0$ and $103.05\Omega_0^{-1}$. The solid lines represent the $\beta$ tilt profiles computed for our turbulent bending wave runs. Meanwhile the corresponding dashed lines denote the inviscid, laminar solutions examined in detail in Section \ref{sec:laminar_results}. Finally, the dotted lines represent similarly set-up hydrodynamic numerical solutions but with a shear viscous $\alpha$ switched on. In order to compare the efficacy of this viscous prescription with the fully turbulent solutions we motivate the choice of $\alpha$ from the volume averaged stress diagnostic computed in Section \ref{subsec:MRI_charactersation}. Therefore for the $L_\alpha$ and $H_\alpha$ runs we adopt $\alpha = 0.0066$ and $\alpha = 0.022$ respectively. Whilst the stress profile in Fig.~\ref{fig:stress_profile} exhibits some radial variation, here we adopt a spatially constant $\alpha$ as a first order approximation. Indeed, in the bulk of the disc away from the buffer zones, the stress profile is relatively uniform  where the bending wave starts evolving.

% -.-.-.-.-.-.-.-.-.-.-.-.-.-.-.-.- %
% -.-.-.-.-.-.-.-.-.-.-.-.-.-.-.-.- %

\begin{figure}
    \centering
    \includegraphics[width=\columnwidth]{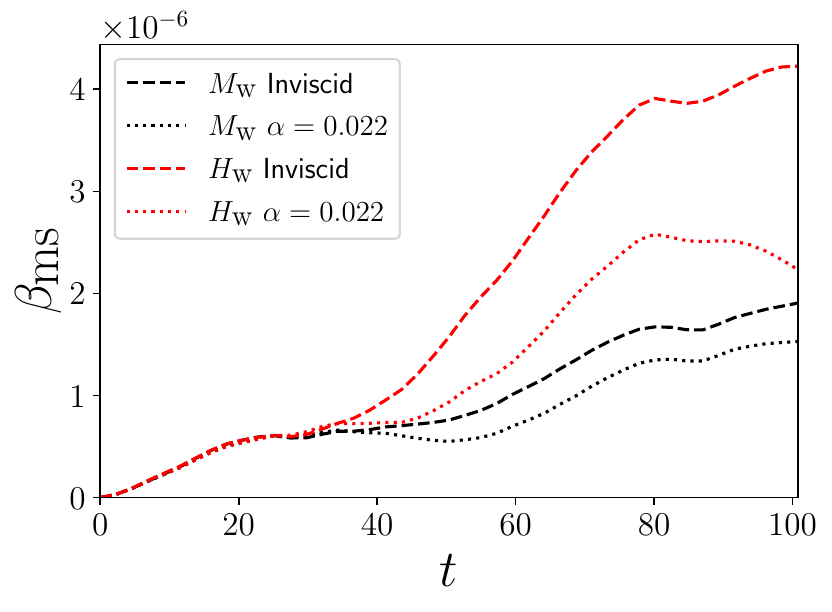}
    \caption{The mean-squared error $\beta_\textrm{ms}$, between the turbulent run tilt profile and that of the inviscid (dashed) or viscous (dotted) laminar runs. We compute this diagnostic for the $(M_\textrm{w},H_\alpha)$ (black) and $(H_\textrm{w},H_\alpha)$ (red) panels shown in Fig.~\ref{fig:turb_bending}.}
    \label{fig:turb_error_high}
\end{figure}

Most notably, for the $L_\textrm{w}$ runs the initial Gaussian tilt is essentially erased immediately by the turbulent velocity field.  Furthermore, the $H_\alpha$ state presents even greater stochasticity in the ensuing noisy tilt profile. This confirms our expectations from panel (a) of Fig.~\ref{fig:turbulent_power} wherein the turbulent velocity overwhelms the characteristic perturbations induced by the bending wave. This is particularly significant since as discussed in Section \ref{sec:laminar_results}, it is these lowest tilts which are most robustly captured by linear theory. Therefore only a weak level of turbulence can significantly disrupt the detailed analytical predictions. In contrast, when one moves towards the higher tilt amplitudes $M_\textrm{w}$ and $H_\textrm{w}$, the qualitative character of the bending wave is preserved. The pulse approximately tracks the profile for the inviscid, laminar bending wave with a noisy residual deviation on top. This scatter is intuitively more pronounced for the $H_\alpha$ runs. The viscous plots are only slightly different to the inviscid runs, exhibiting a vaguely more diffusive character when $\alpha = 0.022$. Nonetheless, these viscous runs are perhaps a better fit for the turbulent propagation as can be seen more quantitatively in Fig.~\ref{fig:turb_error_high}.

% -.-.-.-.-.-.-.-.-.-.-.-.-.-.-.-.- %
% -.-.-.-.-.-.-.-.-.-.-.-.-.-.-.-.- %

Here we extract the mean-squared error $\beta_\textrm{ms}$, computed from the residual difference between the turbulent $\beta$ profile and the inviscid or viscous laminar results (the dashed vs dotted lines). We perform this analysis on the high $\alpha$ state $H_\alpha$, $M_\textrm{w}$ and $H_\textrm{w}$ runs which are coloured black and red respectively. For both the medium and high warp cases considered, the laminar $\alpha = 0.022$ prescription helps reduce the residuals at late times, compared with the inviscid case. Initially all lines overlap and exhibit an approximately linear growth of $\beta_\textrm{ms}$. This is suggestive of a random-walk process whereby the turbulence stochastically kicks the ordered tilt profile towards a more disordered state about the laminar solution. This seems to temporarily plateau at around $t \sim 20\Omega_0^{-1}$. Of course, a random-walk in $\beta$ at any radial location must ultimately be constrained by the conservation of total angular momentum and radial mixing of these perturbations. Indeed, even our initially flat reference MRI disc has some noisy tilt profile with a saturated mean-squared tilt profile, related to the strength of the turbulence. If we instead plot the equivalent $\beta_\textrm{ms}$ curves for the low turbulent state (not shown), we find that the initial growth rate of the mean-squared error is 1/3 as fast and begins levelling off at 1/3 of the high state value. This is in keeping with the findings of Fig.~\ref{fig:stress_profile} which quantifies the turbulence in the high state as being three times stronger compared with the low state. Later on, beyond $t>40\Omega_0^{-1}$, we see that $\beta_\textrm{ms}$ begins to rise again, coincident with the decomposition of the tilt into inwards and outwards propagating pulses. It is during this stage that the discrepancy between the viscous and inviscid laminar predictions grows (see the separation of the dashed and dotted lines in Fig,~\ref{fig:turb_bending} at later times). Similarly, the renewed growth in the $\beta_\textrm{ms}$ tracks suggests that the turbulent evolution is also distinct from the laminar cases here.

% ============================ %
\section{Discussion} 
\label{sec:discussion}
% ============================ %

Distorted geometries are the general outcome of any misalignment possessed by the disc system, and hence understanding their most fundamental manifestation as linear bending waves is extremely important. However, this study has demonstrated that previous analytical and numerical comparisons of simple bending waves are often incomplete and highlighted that there is more to learn from controlled experiments. In this discussion we will examine the consequences of our laminar bending wave results in Section \ref{subsec:discuss_laminar} before considering the implications for the turbulent disc in \ref{subsec:discuss_turbulent} 

% ---------------------------- %
\subsection{Connection of laminar results to previous work}
\label{subsec:discuss_laminar}
% ---------------------------- %

Linear, laminar theory is the most analytically tractable avenue for interpreting warped discs and has therefore received some close study in the past. Originally, \NP99 performed SPH simulations to investigate the propagation of free bending waves in a setup similar to ours. Whilst their focused study captured many of the qualitative features predicted by linear theory, the low number of particles employed results in large numerical viscosities. In order to cleanly extract the bending wave from numerical noise, they initialise larger tilts with amplitude $>5^\circ$. As mentioned in Section \ref{subsec:weak_nonlinearities} this begins to engage weak nonlinearities which will lead to departures from the strictly linear theory. Furthermore, since the particle resolution is proportional to the fluid density, numerical artefacts become significant high above the midplane. This obfuscates a detailed comparison with the internal flow fields. In contrast, we are able to probe very low amplitude bending perturbations which allow for an unprecedented comparison with the theory. Whilst more recent grid-based and higher resolution SPH studies have conducted a similar experiment in order to benchmark their code performance, they do not rigorously interrogate the linear theory they are comparing with. For example, \cite{FragnerNelson_2010}, \cite{NealonEtAl_2015} and \cite{ArzamasskiyEtAl_2018} compare the $g$ variable from the reduced bending wave theory outlined in Section \ref{subsec:reduced_bending}, with the $\beta$ measured from simulations. As pointed out in Section \ref{subsec:comparison_diagnostics} this is an erroneous identification and leads to the underestimation of code performance. Furthermore, there is no examination of the internal flow structure in these previous works, despite this being an essential part of the warped disc dynamics. Our work therefore sets a robust standard for capturing bending waves and underlines grid-based codes as an effective tool for studying distorted discs. This nicely complements the recent study of \cite{KimmigDullemond_2024} who also perform a more comprehensive study of warped disc evolution using grid-based codes. However, they compare larger amplitude distortions with the 1D generalised warping formalism developed in \cite{MartinEtAl_2019} and \cite{DullemondEtAl_2022a}. The assumptions going into this theory somewhat obscures whether discrepancies are due to numerical error or analytical approximation. For example, the development of twist is not predicted by their theory but found in the simulations. The twist found in our numerical experiments agrees well with the Fourier-Hermite formalism (see Fig.~\ref{fig:hydro_linear_comparison}) and hence emphasises the lessons which can still be learned from linear theory and extrapolated towards nonlinear studies. To gain some intuition as to why the twist develops in our reduced or full Fourier-Hermite theory, but not in the warping formalism of \cite{LubowOgilvie_2000} or \cite{DullemondEtAl_2022a}, we can appeal to the local shearing box dispersion relation for $n = 1$ waves of the form $\exp(i k R - \omega t)$ in a Keplerian, isothermal disc \citep[c.f.][]{OkazakiEtAl_1987,Ogilvie_2022},
\begin{equation}
    \label{eq:local_dispersion}
    (\omega^2-\Omega_{\rm K}^2)^2 = c^2 k^2 \omega^2 .
\end{equation}
This local approach requires $|k R| \gg 1$. However, we are often interested in the case where the bending wavelength is longer than the scale height such that $ |k H| = |k R h| < 1$. In this intermediate regime, one can expand the dispersion relation in the small parameter $k H$ to find the dispersion branches \citep[][]{LubowPringle_1993},
\begin{equation}
    \frac{\omega}{\Omega_{\rm K}} = 1 \pm \frac{1}{2} k H - \frac{1}{8}(k H)^2 + \mathcal{O}\left[(k H)^3\right],
\end{equation}
which are visualised in Fig.~1 of \cite{Ogilvie_2022}. The long wavelength theory of \cite{LubowOgilvie_2000} adopts asymptotic scalings with $|k H| \ll 1$ such that only the leading order, non-dispersive term contributes to the wave group velocity at $c/2$. The connection between the general linearization procedure described in Section \ref{sec:theory}, with this long wavelength asymptotic limit is formally demonstrated in appendix \ref{app:long_wavelengths}. In effect, the leading order dominant balance corresponds to a rigid tilting of the disc, wherein the real and imaginary terms in our evolutionary equations cancel identically. This precludes the development of an imaginary phase evolution of $v_z$ and hence the disc cannot twist up. However, for smaller bending wavelengths, the higher order corrections enter and the wave will propagate dispersively. This was investigated in the case of a travelling bending pulse by \cite{Ogilvie_2006}. He found that the dispersive terms cause the warp to evolve according to a Schrödinger-like equation. Just as a real Gaussian wavepacket develops an imaginary wavefunction component, the complex nature of this equation corresponds to an azimuthal phase evolution or twisting of the warp. Whilst this is a purely linear effect, \cite{Ogilvie_2006} also considers the additional dispersive and twisting contributions which arise from nonlinearities, which might also be at play in the simulations of \cite{KimmigDullemond_2024}.

Finally, the simple extension of our reduced theory towards locally isothermal discs (see Section \ref{subsec:reduced_bending}) also offers a straightforward way to test code performance across different thermal profiles. Both \cite{ArzamasskiyEtAl_2018} and \cite{KimmigDullemond_2024} only considered globally isothermal discs. Meanwhile, \cite{FragnerNelson_2010} adopt a disc profile with constant aspect ratio and therefore temperature exponent of $s = 1$. However, they compare this with the formalism developed in \NP99 which is formally valid for globally isothermal discs only.

% ---------------------------- %
\subsection{Turbulent implications}
\label{subsec:discuss_turbulent}
% ---------------------------- %

Extending our results to turbulent discs reveals distinctly different behaviour. In particular, for the lowest tilting regime, where we can most confidently assert linear theory, the turbulent velocity profile rapidly disrupts the propagating bending wave (see Fig.~\ref{fig:turb_bending}). Meanwhile for the larger amplitude tilts, although the qualitative behaviour of the bending wave is preserved for longer, significant scatter in the tilt profile builds up quickly and fitting the evolution with a simple viscous $\alpha$ prescription is perhaps inadequate. Indeed, we have characterised the background MRI turbulent state in detail showing that the velocity spectrum can be comparable to the flows excited by bending waves. Furthermore, the dominant horizontal length scales are larger than $H$ which suggests a local, viscous approach will be unsuitable for pulses with a similarly short radial width. Indeed, $\alpha$ models implicitly assume some scale separation between the fine grained turbulence and the dynamical scales of interest. For asymptotically long wavelength warps (with radial variations $\gg H$) considered in the viscous theories of \cite{Ogilvie_1999,OgilvieLatter2013a}, such an approximation is better justified. Indeed, previous support for warped $\alpha$ models has been advocated by \cite{TorkelssonEtAl_2000} who found that the warped disc sloshing motion, excited in a shearing box, decays according to an isotropic viscous prescription. This is perhaps surprising given the anisotropic turbulent structure emergent in local and global simulations of the MRI \cite[e.g.][]{HawleyEtAl_1995,FlockEtAl_2011}. However, the interplay of the MRI with the vertical sloshing motions is still not well understood. \cite{ParisOgilvie_2018} incorporated strong vertical magnetic fields into the warped shearing box model of \cite{OgilvieLatter2013a}, neglecting the presence of the MRI. Even this `simple' setup leads to interesting dynamics as the magnetic tension provides an additional epicyclic restoring force -- effectively providing a `stiffness' along field lines. Whilst net vertical stresses might therefore be negligible in the flat MRI disc, strong magnetic filaments (see the plasma $\beta_\textrm{m}$ slice in Fig.~\ref{fig:turbulent_snapshot}) might still resist epicylic motions induced by bending waves. Extracting such stresses from global simulations is certainly challenging as the usual orbit averaging necessarily cancels to zero over a full sloshing period.

In this study we have invoked the MRI as our source of turbulence. Of course, in this global simulation we do not expect the nature of the turbulence to be fully converged. In a sense we are agnostic as to the `true' behaviour of the MRI and are instead motivated by the development of turbulence in general. Indeed, a variety of other disc instabilities could also be active in different environments -- stirring the disc in myriad ways. For example the nonlinear state of the \textit{Vertical Shear Instability} (VSI) \citep{NelsonEtAl_2013,BarkerLatter_2015} is characterised by vertical corrugation modes whilst the gravitational instability (GI) \citep{Toomre_1969,KratterLodato_2016} generates shearing spiral waves. Furthermore, distorted discs themselves can trigger the \textit{parametric instability} \cite{GammieEtAl_2000}, which excites small scale inertial waves and may rapidly damp the warp \citep{DengEtAl_2020,FairbairnOgilvie_2023}. Considering the distinct effects of each turbulent mechanism in their respective environments will therefore prove an interesting avenue for further study.

Accepting the complications of disc turbulence, there are several practical consequences that merit future investigation. For example, the excitation and propagation of bending waves mediates the interaction between an inclined planet and the surrounding disc \citep{TanakaWard_2004,ArzamasskiyEtAl_2018}. Previous studies have instead focused on the coplanar interaction between a perturber and a turbulent disc \cite[e.g.][]{BaruteauLin_2010,StollEtAl_2017,WuEtAl_2024}. They find that the stochastic turbulence can lead to profound differences in the migration behaviour when the mixing is strong enough to disrupt the classical, laminar wake structure. Similarly, we might anticipate the properties of inclination damping to change when the wave response is significantly modified by turbulence. This could affect the alignment architectures of planetary systems \citep{BitschKley_2011,MillhollandEtAl_2021} or the capture rate of black holes into binaries in Active Galactic Nuclei (AGN) \citep{TagawaEtAl_2020,WangEtAl_2024}.

% ---------------------------- %
\subsection{Extension to nonlinear warps}
\label{subsec:nonlinear_warps}
% ---------------------------- %

In this study we have focused on a detailed understanding of bending waves with small tilt amplitudes and therefore subsonic velocity perturbations. The picture can change markedly as one transitions towards nonlinear warps. Whilst a theory for large wavelength, nonlinear warps has been developed by \cite{Ogilvie_1999,ParisOgilvie_2018}, this is not applicable to the shorter wavelength pulses considered here. Instead, previous numerical studies have probed this regime with \NP99 finding that the short radial variation in tilt leads to counter propagating flows which shock, dissipate energy and efficiently diffuse the warp when $R |d \beta/ dR| > h$. Similar behaviour was found by \cite{SorathiaEtAl_2013a} wherein large warps over finite radial widths induce transonic flows which shock and mix angular momentum. This was subsequently extended towards magnetised turbulence in tilted tori around spinning black holes, where the disc warping facilitates Bardeen-Petterson alignment \citep{SorathiaEtAl_2013b}. The Lense-Thirring torques drive nonlinear warp amplitudes which once again exhibit pressure dominated dynamics. Nonetheless, the turbulence leads to some key differences. The stirring of the disc promotes enhanced angular momentum cancellation, even when the warp has decayed below nonlinear amplitudes, thus allowing it to flatten more effectively than a purely hydrodynamic disc. Furthermore, the disruption of bending waves (as seen clearly in Fig.~\ref{fig:turb_bending}) diminishes the radial communication in the disc and hence inhibits solid body precession. This allows larger warped distortions to be established by the Lense-Thirring torque which accelerates the alignment process. 

In fact, as one increases the initial misalignment between the disc orientation and black hole spin, the disc can dramatically \textit{break} apart into separated annuli \citep{NixonKing_2012,NixonEtAl_2012}. This \textit{breaking} or \textit{tearing} process can similarly occur in circumbinary discs which are tilted with respect to the plane of the binary. This fundamentally nonlinear process has been found in a variety of SPH \citep[e.g.][]{NealonEtAl_2016,YoungEtAl_2023} and grid based viscous simulations \citep{RabagoEtAl_2024}, with theoretical efforts to interpret it relying on isotropic $\alpha$ models of warp evolution \citep[][]{DoganEtAl_2018,DoganNixon_2020}. Significantly, recent MRI turbulent simulations have also found disc breaking to occur \citep{KaazEtAl_2023}. The anisotropic directionality and finite length scale of turbulence, exemplified in our work, might change the nature of warp communication over the short length scales associated with the breaking location. Therefore, understanding breaking criteria in the presence of turbulence is a challenging problem which deserves further investigation.
% ============================ %
\section{Conclusions} 
\label{sec:conclusion}
% ============================ %

In this work we have revisited the propagation of bending waves in unprecedented detail, combining theoretical predictions with numerical simulations to gain new insight into their laminar and turbulent evolution. 

We have carefully expounded the linearized theory using a Fourier-Hermite formalism, which naturally simplifies to the reduced bending wave theory of \cite{NelsonPapaloizou1999} when restricted to the lowest order modes. This easily allows us to generalise their equations towards locally isothermal discs. This theory has been rigorously bench-marked with global, 3D grid-based simulations which exhibit remarkable agreement with the warp evolution as well as the detailed internal flow dynamics. This marks a more focused investigation which improves on similar bending wave tests conducted previously. These studies did not explore the comparison with theory as comprehensively -- typically employing more numerically diffusive setups or making erroneous identifications with the theory. In particular our work highlights the need to incorporate higher order polynomial modes to accurately capture the twisting profile and velocity structure of bending waves, thus setting a new standard for bending wave code performance. We find that this linear theory continues to describe our simulations well even as the tilt amplitude is boosted towards the weakly nonlinear regime, despite density perturbations becoming of order unity. This is caveated with the subtlety that the actual initial condition of the warped disc should only be projected onto the odd vertical modes, since the even modes are tempered by nonlinearities which are neglected in the linearized equations. 

This laminar investigation is complemented with a study of bending waves in turbulent discs. We establish a magnetised turbulence via the toroidal MRI which exhibits high and low states of stirring, quantified via effective $\alpha$ stress parameters. This simple diagnostic belies the more complex and anisotropic turbulence which is described in some detail. We find that the velocity perturbations and characteristic eddy length scales are effectively able to disrupt small amplitude bending waves. Whilst larger amplitude bending waves can persist qualitatively in character, they become significantly modified due to the interaction with the turbulence. The use of a viscous $\alpha$ model to describe this behaviour is visibly unsatisfactory for these short wavelength bending waves, but might still be suitable for longer wavelength warps where the dynamical length scales are well separated from the scale of turbulence. Investigating how turbulent bending waves affect inclined planet-disc interactions, AGN black hole binary assembly rates and disc breaking criterion are all fruitful avenues for further research.

%% Also note that the akcnowlodgment environment does not support long amounts of text. If you have a lot of people and institutions to acknowledge, do not use this command. Instead, create a new \section{Acknowledgments}.

\begin{acknowledgments}
C.W.F would like to thank the anonymous referee for useful comments and suggestions which helped improve this manuscript. C.W.F also acknowledges helpful conversations with Jim Stone, Chris White, George Wong and Lizhong Zhang which often inspired solutions to numerical problems. Furthermore, discussions with Gordon Ogilvie helped clarify subtleties of bending wave theory. This work was supported by funding from the W. M. Keck Foundation Fund, the Adler Family Fund, and the Sivian Fund at the Institute for Advanced Study.
\end{acknowledgments}


\vspace{5mm}
% \facilities{}


\software{\texttt{Athena++} \citep{StoneEtAl_2019}}



\appendix
% ........................................ %
\section{Viscous bending waves}
\label{app:viscous}
% ........................................ %

In our study of laminar bending waves carried out in Section \ref{sec:laminar_results} we focused on the inviscid regime with $\alpha = 0$ and ignored the viscous stress tensor given by equation \eqref{eqn:athena_viscous}. We will now briefly discuss how the theory and simulations change in the presence of viscosity. This is immediately complicated by the fact that a non-zero $\alpha$ induces an accretion flow which precludes a simple equilibrium structure on which to perform our linear analysis. However, \cite{PapaloizouLin_1995} suggest that if this background accretion can be ignored, then the dominant viscous effect is the damping of the warp induced, horizontal sloshing motions via the accelerations
\begin{equation}
    a_R = \frac{\partial}{\partial z}\left(\rho_\textrm{td} \nu \frac{\partial u_R^\prime}{\partial z}\right) \quad \textrm{and}\quad a_\phi=\frac{\partial}{\partial z}\left(\rho_\textrm{td} \nu \frac{\partial u_\phi^\prime}{\partial z}\right),
\end{equation}
which should be appended to the radial and azimuthal components of equation \eqref{eq:euler_perturbation}. Performing the usual Fourier-Hermite decomposition on these terms yields 
\begin{equation}
    \label{eq:visc_horizontal}
    -\alpha \Omega_\textrm{K} n u_{R,n}^\prime \quad \textrm{and} \quad -\alpha \Omega_\textrm{K} n u_{\phi,n}^\prime,
\end{equation}
which are added to the right hand sides of equations \eqref{eq:fourier_hermite_uR} and \eqref{eq:fourier_hermite_uphi} respectively. This approach has previously been employed by \NP99 when benchmarking their SPH simulations of bending waves against linear theory. Unfortunately, the difficulty in constraining SPH numerical viscosity prevents a controlled comparison. Furthermore, the assumption that the induced background accretion is negligible has not been interrogated sufficiently in this context.

Indeed, lets presently consider a globally isothermal disc. The action of viscosity on the background, axisymmetric equilibrium described in section \ref{subsec:background} produces the standard accretion stress component which acts on the radially shearing azimuthal flow,
\begin{equation}
    \Pi_{R \phi} = \rho_\textrm{td}\nu R\frac{\partial \Omega_\textrm{g}}{\partial R} = -\rho \nu q\Omega_\textrm{K}\left[1+\mathcal{O}(h^2)\right],
\end{equation}
where $q = 3/2$ is the Keplerian shear parameter and for simplicity we will subsequently drop the $\mathcal{O}(h^2)$ contributions. The divergence operator acting on this term gives the axisymmetric background forcing which is tacked on to the $\phi$ component of equation \eqref{eq:euler_perturbation} as $\left(\partial \Pi_{R\phi}/\partial R+2\Pi_{R\phi}/R\right)/\rho$. In this sense, the action of viscosity on the inviscid equilibrium drives a steady forcing which takes the disc away from this equilibrium. We once again project this forcing onto the Fourier-Hermite basis, which yields additional accelerations 
\begin{align}
    & \frac{\partial u_{\phi,0}^\prime}{\partial t} = -\alpha c h q \Omega_\textrm{K} (2+\Gamma+\chi-q), \nonumber \\
    \label{eq:visc_background}
    & \frac{\partial u_{\phi,2}^\prime}{\partial t} = -\alpha c h q \Omega_\textrm{K} \chi ,
\end{align}
which are to be added to the right hand sides of the $n=0$ and $n=2$ components of $m=0$ axisymmetric azimuthal perturbation equations.

One can then ask whether the effects of the damping \eqref{eq:visc_horizontal} or background accretion \eqref{eq:visc_background} are more important. From equation \eqref{eq:reduced_tilt_twist} we see that the typical velocity perturbations excited by bending waves are of the order $u^\prime \sim R\Omega \beta$. Then the characteristic timescale for these sloshing motions to damp via equations \eqref{eq:visc_horizontal} is given by $t_\textrm{damp} \sim \Omega^{-1}/\alpha$. This competes with the timescale for the background accretion to induce comparable flows, $t_\textrm{bg} \sim \beta\Omega^{-1}/(\alpha h^2)$. Therefore, we expect the background viscous term will become important relative to the damping when $\beta \lesssim h^2$ which is satisfied for the smallest, linear tilts with disc thicknesses used in our simulations. Indeed, for the smallest warps considered here $\beta_\textrm{amp} = 0.00175 < h(r_\textrm{mid})^2 = 0.0064$. 
%
% -.-.-.-.-.-.-.-.-.-.-.-.-.-.-.-.- %
\begin{figure*}
    \centering
    \includegraphics[width=\textwidth]{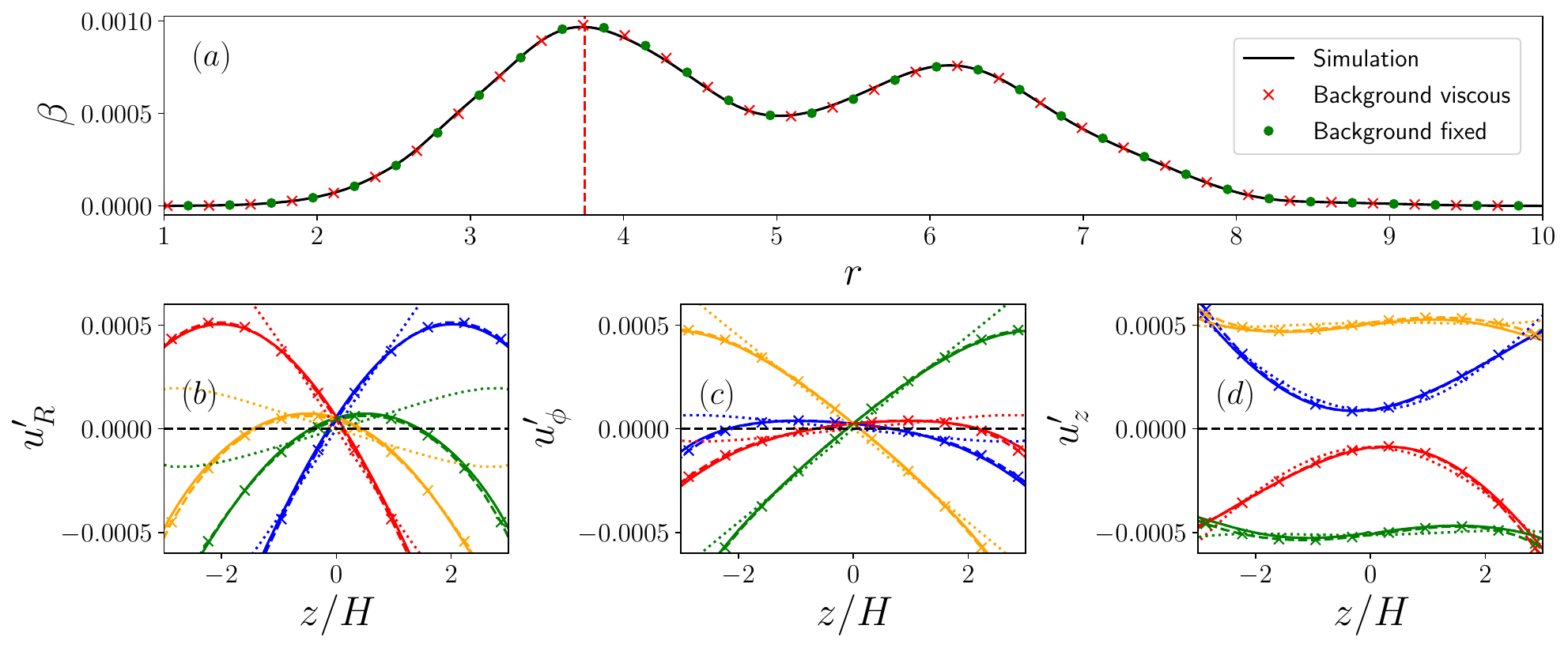}
    \label{fig:viscous_bending}
    \caption{Viscous evolution for a bending wave with $\alpha = 0.005$ and $\beta_\textrm{amp} = 0.00175$. (a) $\beta$ profile at $t = 68.7\Omega_0^{-1}$ for the simulation (solid black line). The markers denote linear calculations which include viscous damping of horizontal motions \eqref{eq:visc_horizontal}. However, the red crosses also include the background accretion terms \eqref{eq:visc_background} whilst the green dots exclude it. Panels (b), (c) and (d) show the velocity perturbations ($u_R^\prime$,$u_\phi^\prime$,$u_z^\prime$) through the cylindrical radial slice at $R = 3.75$, marked by the red dashed line in panel (a). Simulation results = solid lines, linear theory without background accretion = dotted lines and with background accretion = crosses. The colours (blue, yellow, red, green) denote 4 azimuthal phases $\phi = [0,\pi/2,\pi,3\pi/2]$ respectively.}
\end{figure*}
% -.-.-.-.-.-.-.-.-.-.-.-.-.-.-.-.- %
%
The importance of the background accretion is clearly demonstrated in Fig.~\ref{fig:viscous_bending} where we examine the tilt evolution and internal flows in a viscous disc with $\alpha = 0.005$ and $\beta_\textrm{amp} = 0.00175$. We simply run the globally isothermal simulation as before, but now switch on the viscous shear term and compare this with the Fourier-Hermite linear theory, supplemented with the additional terms discussed above. Since these new terms force the $m=0$ and $n = (0,2)$ equations, we now include the $m = [0,1,2]$ and $n = [0,1,2,3,4,5]$ modes in our solution. In panel (a) we plot the tilt evolution at $t = 68.7\Omega_0^{-1}$. The solid line from the simulation is closely traced by the linear theory predictions which are in fact very similar to the inviscid run\footnote{We actually find that notable differences compared with the inviscid tilt profile only emerge when the disc transitions towards the diffusive regime with $\alpha \gtrsim h$}. Whilst the green dots only include the additional viscous terms from the horizontal damping, described  by equations \eqref{eq:visc_horizontal}, the red crosses also incorporate the background accretion terms given by \eqref{eq:visc_background}. Since the background accretion only drives axisymmetric, even vertical modes, these have little effect on the integrated tilt profile for such low values of $\alpha$. However, we clearly see their effect on the internal flow fields for $(u_R^\prime,u_\phi^\prime,u_z^\prime)$ shown in panels (b), (c) and (d) for the radial slice at $R=3.75$ at 4 azimuthal phases. The solid lines once again show the simulation results whilst the dotted lines represent the linear theory excluding the background accretion. This shows a notable discrepancy away from the midplane. In fact, when the background accretion is included, the linear theory performs much better with the crosses nicely overlying the simulation results. In particular, including these even $n$ modes notably changes the vertical flow parity, so now the horizontal sloshing modes clearly deviate from being linear coordinates of $z$.

This foray into viscous discs acts as a cautionary warning for the neglect of the accretion term in this controlled problem. Indeed, for the smallest amplitude warps, where we can most robustly test linear theory, the background accretion becomes important as the system evolves away from the inviscid steady state. Whilst we have taken $\beta \ll h$ to illustrate this effect, the linear theory of \cite{LubowOgilvie_2000} assumes $\beta \sim h $ which allows the background accretion to be formally separated from the warp evolution. However, even for larger tilts, where the viscous damping terms are more important, they will only notably affect the bending wave over the radial propagation provided $t_\textrm{damp}$ is less than the characteristic bending wave crossing time, $R/c = \Omega^{-1}/h$. This is is only satisfied when $\alpha > h$ which suggests a transition towards the diffusive regime of warp evolution, which we do not explore  in this paper. 

Finally, it should be acknowledged that in this discussion we have only considered the damping due to the vertical shearing of horizontal sloshing motions. However, for the relatively short wavelength tilts here, other components of the perturbed viscous stress might become important also. This would require a full projection of the stress tensor onto the Fourier-Hermite basis which is, from experience, a tedious but doable process. Such a procedure does not provide much physical insight however and instead we defer such short wavelength, diffusive tilt evolution to numerical simulation.

% ........................................ %
\section{Connection with long bending wavelength theory}
\label{app:long_wavelengths}
% ........................................ %

The general linearization procedure described in Section \ref{sec:theory} is found to present fantastic agreement with the numerical experiments of short wavelength bending waves. This theory fully captures the detailed warp evolution, including the development of twist, compared with the long wavelength theories of \cite{LubowOgilvie_2000} and \cite{DullemondEtAl_2022a}. Whilst these other approaches sacrifice some degrees of freedom, this compromise effectively recasts the evolutionary equations in a more physically intuitive form. In this section I will demonstrate how the general linear theory connects to these equations in the long wavelength, asymptotic limit.

For illustration purposes, consider the simplest case of a globally isothermal disc ($s=0$) with a uniform midplane density profile ($d=0$) such that $\Omega = \Omega_{\rm K}$ and the scale height flares according to $\chi = 3/2$. Furthermore, restrict the full set of Fourier-Hermite modes in equations \eqref{eq:fourier_hermite_uR}-\eqref{eq:fourier_hermite_W} to the reduced warping set with $m = n = 1$, as discussed in Section \ref{subsec:reduced_bending}. In the long bending wavelength limit we can exploit the large scale separation between the radial and vertical dimensions where we choose $R\sim\mathcal{O}(1)$ and $H\sim\mathcal{O}(\epsilon)$, where $\epsilon$ is a small parameter characterising of the thin disc aspect ratio. Specifically we will scale $\Tilde{H} = H/\epsilon$ to be of order unity. Motivated by the slow bending wave evolution over the sound speed crossing time, the length-scale separation permits the introduction of a slow time-scale $T = \epsilon t$ over which the warp geometry changes. We can then introduce the asymptotic expansion
\begin{align}
    & u_{R,1}^\prime = \epsilon u_{R}^{(1)}(R,T)+\epsilon^2 u_{R}^{(2)}(R,T)+\mathcal{O}(\epsilon^3) ,  \\
    & u_{\phi,1}^\prime = \epsilon u_{\phi}^{(1)}(R,T)+\epsilon^2 u_{\phi}^{(2)}(R,T)+\mathcal{O}(\epsilon^3) ,  \\
    & u_{z,1}^\prime = \epsilon u_{z}^{(1)}(R,T)+\epsilon^2 u_{z}^{(2)}(R,T)+\mathcal{O}(\epsilon^3) ,  \\
     & W_{1} = \epsilon^2 W^{(1)}(R,T)+\epsilon^3 W^{(2)}(R,T)+\mathcal{O}(\epsilon^4).
\end{align}
These are inserted into \eqref{eq:fourier_hermite_uR}-\eqref{eq:fourier_hermite_W} and arranged into a hierarchy of equations in terms of powers of $\epsilon$ which are solved successively. At leading order $\mathcal{O}(\epsilon^1)$, we obtain
\begin{align}
    \label{eq:epsilon_1_uR}
    & i \Omega_{\rm K} u_{R}^{(1)}-2\Omega_{\rm K} u_{\phi}^{(1)} = 0 , \\
    \label{eq:epsilon_1_uphi}
    & i \Omega_{\rm K} u_{\phi}^{(1)} + \Omega_{\rm K} u_{R}^{(1)}/2 = 0 , \\
    \label{eq:epsilon_1_uz}
    & i \Omega_{\rm K} u_{z}^{(1)} = - W^{(1)}/ \Tilde{H} .
\end{align}
The general epicylic oscillation solution to equations \eqref{eq:epsilon_1_uR} and \eqref{eq:epsilon_1_uphi} is clearly 
\begin{equation}
    \label{eq:epsilon_1_epicycle}
    u_{R}^{(1)} = U(R,T) \quad \textrm{and} \quad u_{\phi}^{(1)} = \frac{1}{2}i U(R,T).
\end{equation}
Meanwhile, the $u_{z,1}^\prime$ equation is consistent with the rigid tilting of disc annuli such that 
\begin{equation}
\label{eq:complex_tilt}
\mathcal{W}(R,T) \equiv -\frac{i u_{z}^{(1)}}{R\Omega_{\rm K}} = \frac{(L_x+i L_y)}{\epsilon |\mathbf{L}|} ,
\end{equation}
encodes the horizontal projection of the unit tilt vector at each radius. At next order $\mathcal{O}(\epsilon^2)$, the $u_{R,1}^\prime$ and $u_{\phi,1}^\prime$ equations yield
\begin{align}
    \label{eq:epsilon_2_uR}
    & \frac{\partial u_{R}^{(1)}}{\partial T} + i \Omega_{\rm K} u_{R}^{(2)}-2\Omega_{\rm K}u_{\phi}^{(2)} = -\frac{\partial W^{(1)}}{\partial R}+\frac{\chi}{R}W^{(1)}, \\
    \label{eq:epsilon_2_uphi}
    & \frac{\partial u_{\phi}^{(1)}}{\partial T} + i \Omega_{\rm K} u_{\phi}^{(2)}+\frac{1}{2}\Omega_{\rm K} u_{R}^{(2)} = -\frac{i}{R}W^{(1)},
\end{align}
whilst the $u_{z,1}^\prime$ equation provides
\begin{equation}
    \label{eq:epsilon_2_uz}
    \frac{\partial u_{z}^{(1)}}{\partial T}+i \Omega_{\rm K}u_{z}^{(2)}+\frac{W^{(2)}}{\Tilde{H}} = 0.
\end{equation}
Finally we will require the $W_{1}$ equation at $\mathcal{O}(\epsilon^3)$ which gives
\begin{align}
\label{eq:epsilon_3_W}
& \frac{\partial W^{(1)}}{\partial T}+i\Omega_{\rm K} W^{(2)}-\Omega_{\rm K}^2 \Tilde{H} u_z^{(2)} = \nonumber \\
& -\Omega_{\rm K}^2 \Tilde{H}^2 \left[ \frac{(2\chi+1)}{R}u_{R}^{(1)}+\frac{\partial u_{R}^{(1)}}{\partial R}+\frac{i u_{\phi}^{(1)}}{R}\right].
\end{align}
The epicyclic balance which appears at leading order is inherited by the higher order terms in equations \eqref{eq:epsilon_2_uR} and \eqref{eq:epsilon_2_uphi}. Therefore a linear combination of these can eliminate the $u_R^{(2)}$ and $u_\phi^{(2)}$ dependencies leaving the solvability condition
\begin{equation}
    \label{eq:solvability_A}
    \frac{\partial u_{R}^{(1)}}{\partial T} - 2i \frac{\partial u_{\phi}^{(1)}}{\partial T} = -\frac{\partial W^{(1)}}{\partial R}+\frac{\chi}{R}W^{(1)}-\frac{2}{R}W^{(1)} . 
\end{equation}
Similarly one can combine equations \eqref{eq:epsilon_2_uz} and \eqref{eq:epsilon_3_W} to remove the $u_z^{(2)}$ and $W^{(2)}$ dependencies yielding
\begin{align}
    \label{eq:solvability_B}
    & -i \Omega_{\rm K} \Tilde{H} \frac{\partial u_z^{(1)}}{\partial T} +\frac{\partial W^{(1)}}{\partial T} = \nonumber \\
    & -\Omega_{\rm K}^2 \Tilde{H}^2\left[\frac{(2\chi+1)}{R}u_R^{(1)}+\frac{\partial u_R^{(1)}}{\partial R}+\frac{i u_\phi^{(1)}}{R}\right].
\end{align}
We can now insert the leading order solutions given by equations \eqref{eq:epsilon_1_uR}-\eqref{eq:epsilon_1_uz}, alongside the epicylic amplitude and tilt variables given by equations \eqref{eq:epsilon_1_epicycle} and \eqref{eq:complex_tilt} respectively, into our solvability conditions \eqref{eq:solvability_A}-\eqref{eq:solvability_B}. After some manipulation, one can arrange these into a physically intuitive form given by the pair of coupled partial differential equations
\begin{equation}
    \frac{\partial \mathcal{G}}{\partial t} = \frac{\sigma H^2 \Omega_{\rm K}^3 R^3}{4}\frac{\partial \mathcal{W}}{\partial R},
\end{equation}
\begin{equation}
    \sigma \Omega_{\rm K} R^2 \frac{\partial \mathcal{W}}{\partial t} = \frac{1}{R}\frac{\partial \mathcal{G}}{\partial R},
\end{equation}
where 
\begin{equation}
    \label{eq:internal_torque}
    \mathcal{G} = \frac{\sigma H R^2 \Omega_{\rm K}}{2} U
\end{equation}
is to be interpreted as the internal torque induced by the sloshing motions which are excited by the warped geometry. These equations are formally identical to equations (8) and (9) of \cite{LubowOgilvie_2000} in the inviscid, Keplerian limit.

% ........................................ %
\section{Lower resolution comparison}
\label{app:lower_res}
% ........................................ %

% -.-.-.-.-.-.-.-.-.-.-.-.-.-.-.-.- %
\begin{figure*}
    \centering
    \includegraphics[width=\textwidth]{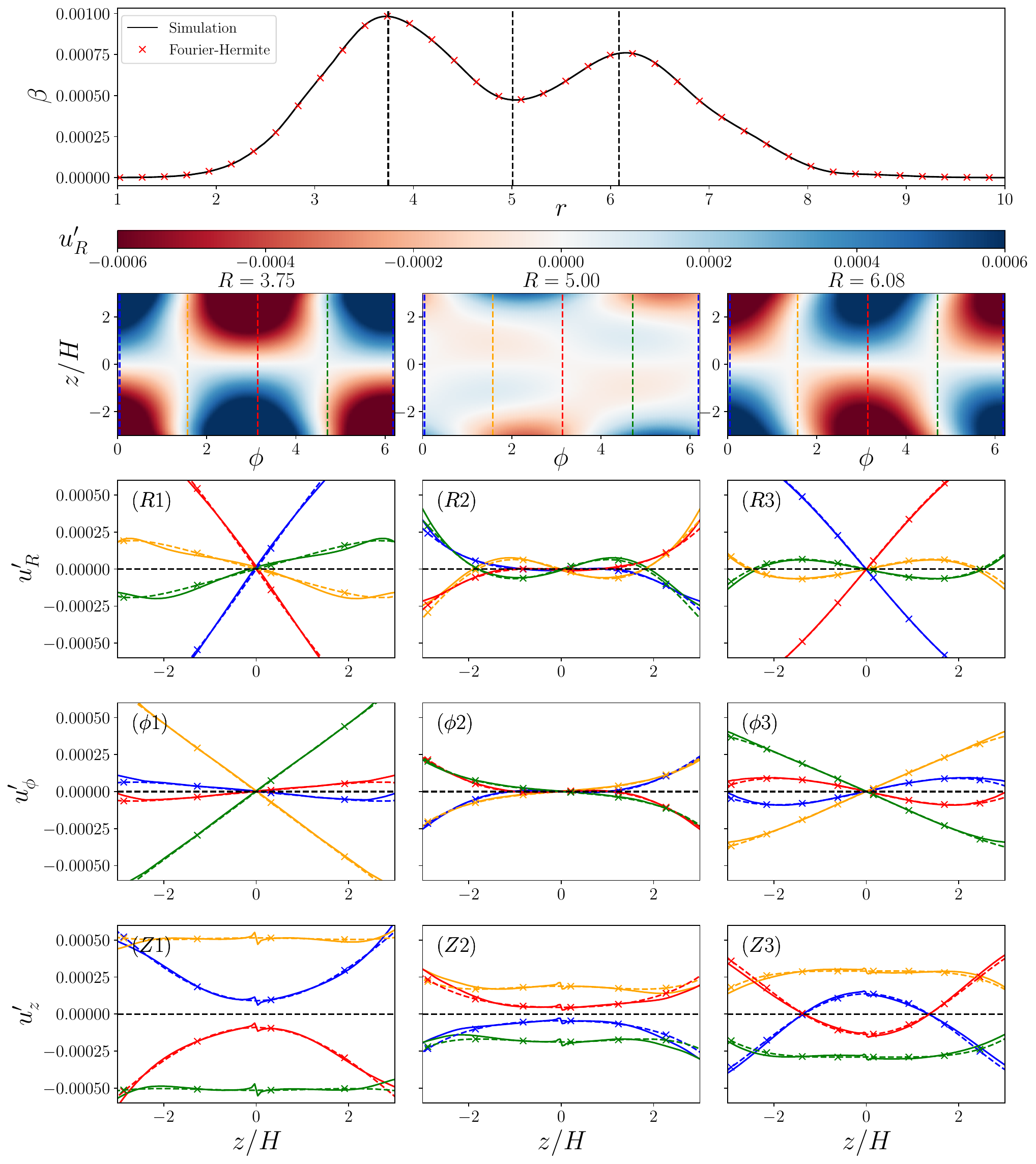}
    \caption{Same as for Fig.~\ref{fig:global_iso_hydro_hermite_panels_0.00175} but for the half resolution run. Note that we don’t consider the point mass potential experiment in this case.}
    \label{fig:global_iso_hydro_hermite_panels_0.00175_half_res}
\end{figure*}
% -.-.-.-.-.-.-.-.-.-.-.-.-.-.-.-.- %

Whilst we will not present a comprehensive resolution study in this paper, here we briefly discuss a half-resolution run of the fiducial globally isothermal, inviscid bending wave, previously examined in Section \ref{sec:laminar_results}. The setup is the same as before, except now we adopt $N_r=256$, $N_\theta=144$ and $N_\phi=512$. The results are shown in Fig.~\ref{fig:global_iso_hydro_hermite_panels_0.00175_half_res}, which performs the same visualization of the warp evolution and internal flows as per Fig.~\ref{fig:global_iso_hydro_hermite_panels_0.00175}. In the upper panel, we still find fantastic agreement for the tilt amplitude $\beta$ across the radial disc domain when comparing the simulation results with the Fourier-Hermite theory. Below this, the internal flow structures also appear to be broadly the same as found in the higher resolution run. Namely, the solid numerical lines closely track the dash-crossed theoretical predictions in all panels. However, there are now subtle deviations which are not present in the higher resolution experiment. Most notably, in the Z panels, there appears to be small discontinuities in $u_z^\prime$ near the midplane, associated with the compressive vertical velocities. Whilst these slight artefacts promote the use of the higher resolution standard employed throughout the paper, the general agreement found here suggests our study is suitably converged.


\bibliography{main}{}
% \bibliography{references}{}
\bibliographystyle{aasjournal}



\end{document}